\def\mode{1} 

\if 0\mode
  \documentclass[manuscript,screen,review]{acmart}
\else
  \documentclass[manuscript,nonacm,screen,
    natbib=false]{acmart}
  \renewcommand\keywordsname{\textbf{Keywords}}
\fi

\usepackage{makecell}
\usepackage{multirow}
\usepackage{hhline}
\usepackage{url}
\usepackage{tikz}
\usetikzlibrary{patterns}
\usepackage{adjustbox}
\usepackage{tocbasic}
\usepackage{lipsum}
\usepackage{framed}
\usepackage{pdflscape} 
\usepackage{afterpage}
\usepackage{booktabs}
\usepackage{longtable}
\usepackage{mdframed}
\usepackage{pgfplots}
\usepackage{pifont}
\usepackage{enumitem}
\pgfplotsset{compat=1.18}

\if 1\mode
  \input{biblatex-config.cls}
\fi

\newcommand{\tocfontsize}{\small}
\expandafter\newcommand\csname r@tocindent4\endcsname{4in}
\DeclareTOCStyleEntries[indent=1ex,pagenumberformat=\tocfontsize,entryformat=\bfseries, entrynumberformat=\tocfontsize,linefill=\tocfontsize\TOCLineLeaderFill,beforeskip=5pt plus 1pt]{tocline}{section}
\DeclareTOCStyleEntries[indent=5ex,pagenumberformat=\tocfontsize,linefill=\tocfontsize\TOCLineLeaderFill]{tocline}{subsection}
\DeclareTOCStyleEntries[indent=10ex,pagenumberformat=\tocfontsize,linefill=\tocfontsize\TOCLineLeaderFill]{tocline}{subsubsection}
\DeclareTOCStyleEntries[indent=15ex,pagenumberformat=\tocfontsize,linefill=\tocfontsize\TOCLineLeaderFill]{tocline}{paragraph}
\DeclareTOCStyleEntries[indent=20ex,pagenumberformat=\tocfontsize,linefill=\tocfontsize\TOCLineLeaderFill]{tocline}{subparagraph}

\AtBeginDocument{%
  }

\setcopyright{acmlicensed}
\copyrightyear{2024}
\acmYear{2024}
\acmDOI{XXXXXXX.XXXXXXX}

\acmConference[]{}{}{}
\acmBooktitle{}
\acmISBN{978-1-4503-XXXX-X/18/06}

\newcommand{\vt}[1]{{\color{purple}\textbf{VT:} #1 $\blacksquare$}}

\newcommand{\lb}[1]{{\color{brown}\textbf{LB:} #1 $\blacksquare$}}

\newcommand{\later}[1]{}

\newcommand{\ie}{\textit{i.e.}~}
\newcommand{\eg}{\textit{e.g.}~}
\newcommand{\token}[1]{\texttt{<#1>}}
\newcommand{\enctype}[2]{\texttt{<#1>} \textit{(#2)}}

\newcommand{\categoryrule}{\midrule[.8pt]}
\newcommand{\headtoprule}{\toprule[1.3pt]}
\newcommand{\modelauthor}[2]{\makecell{\emph{#1}\\#2}}
\newcommand{\Tf}{Transformer}

\newcommand{\sepsubparagraph}{\phantomsection} %

\newcommand{\tableSectionGlobal}[2]{
    \multicolumn{#1}{l}{\raisebox{.1em}{\bfseries #2}} \\
    \categoryrule
}
\newcommand{\tableSection}[1]{}

\newcommand{\availablemodel}[1]{\href{#1}{\ding{51}}}
\newcommand{\notavailablemodel}[1]{\ding{55}}

\makeatletter
\renewcommand\subparagraph{\@startsection {subparagraph}{5}{\z@}{\z@}{\z@}{\hspace{\parindent}\bfseries}}
\makeatother

\makeatletter 
\newcommand\nobreakpar{\par\nobreak\@afterheading} 
\makeatother

\def\oldsubsubsection{} \let\oldsubsubsection=\subsubsection
\renewcommand{\subsubsection}[1]{\oldsubsubsection{#1}\mbox{}\nobreakpar}

\def\oldparagraph{} \let\oldparagraph=\paragraph
\renewcommand{\paragraph}[1]{\oldparagraph{#1}\mbox{}\nobreakpar}

\def\oldsubparagraph{} \let\oldsubparagraph=\subparagraph
\renewcommand{\subparagraph}[1]{\oldsubparagraph{#1}~}

\begin{document}

\title[NLP Methods for Symbolic Music Generation and Information Retrieval: a Survey]
  {Natural Language Processing Methods for Symbolic Music Generation and Information Retrieval: a Survey}






\author{Dinh-Viet-Toan Le}
\authornote{email: \href{mailto:dinhviettoan.le@univ-lille.fr}{dinhviettoan.le@univ-lille.fr}}
\email{dinhviettoan.le@univ-lille.fr}
\orcid{0000-0001-6991-4079}
\affiliation{%
  \institution{Univ. Lille, CNRS, Inria, Centrale Lille, UMR 9189 CRIStAL, F-59000 Lille}
  \city{Lille}
  \country{France}
}

\author{Louis Bigo}
\email{louis.bigo@u-bordeaux.fr}
\orcid{0000-0002-9865-2861}
\affiliation{%
  \institution{Univ. Bordeaux, CNRS, Bordeaux INP, LaBRI, UMR 5800, F-33400 Talence}
  \city{Bordeaux}
  \country{France}}

\author{Mikaela Keller}
\email{mikaela.keller@univ-lille.fr}
\orcid{1234-5678-9012}
\affiliation{%
  \institution{Univ. Lille, CNRS, Inria, Centrale Lille, UMR 9189 CRIStAL, F-59000 Lille}
  \city{Lille}
  \country{France}}

\author{Dorien Herremans}
\orcid{0000-0001-8607-1640}
\affiliation{%
  \institution{Singapore University of Technology and Design}
  \city{Singapore}
  \country{Singapore}}
\email{dorien_herremans@sutd.edu.sg}


\addtocontents{toc}{\protect\setcounter{tocdepth}{-1}}
\begin{abstract}
  \textbf{Abstract --}
  Several adaptations of Transformers models have been developed in various domains since its breakthrough in Natural Language Processing (NLP). This trend has spread into the field of Music Information Retrieval (MIR), including studies processing music data. 
However, the practice of leveraging NLP tools for symbolic music data is not novel in MIR. Music has been frequently compared to language, as they share several similarities, including sequential representations of text and music. These analogies are also reflected through similar tasks in MIR and NLP.

This survey reviews NLP methods applied to symbolic music generation and information retrieval studies following two axes.
We first propose an overview of representations of symbolic music adapted from natural language sequential representations. 
Such representations are designed by considering the specificities of symbolic music.
These representations are then processed by models.
Such models, possibly originally developed for text and adapted for symbolic music, are trained on various tasks. %
We describe these models, in particular deep learning models, through different prisms, highlighting music-specialized mechanisms.
We finally present a discussion surrounding the effective use of NLP tools for symbolic music data.
This includes technical issues regarding NLP methods and fundamental differences between text and music, which may open several doors for further research into more effectively adapting NLP tools to symbolic MIR.

\end{abstract}

\ccsdesc[500]{Computing methodologies~Artificial intelligence}
\ccsdesc[300]{Applied computing~Sound and music computing}
\ccsdesc[300]{Information systems~Music retrieval}

\keywords{
  Music Information Retrieval,
  Natural Language Processing,
  Symbolic music,
  Music generation,
  Music analysis,
  Deep learning.
}


\maketitle

\if 1\mode
{
  \setlength{\parskip}{.3em}
  \hypersetup{hidelinks}
  \tocfontsize
  \addtocontents{toc}{\protect\setcounter{tocdepth}{-1}}
  \tableofcontents
  \addtocontents{toc}{\protect\setcounter{tocdepth}{5}}
}
\pagebreak
\fi

\section{Introduction}
\label{sec:introduction}

The evolution of Natural Language Processing (NLP) has been marked by a substantial journey, progressing from rudimentary rule-based systems like ELIZA~\cite{weizenbaum1966eliza} in 1966 to the widespread adoption of sophisticated deep learning models by the general public, such as ChatGPT. 
In parallel with these advancements, computational music research has been adapting NLP methods for musical data for various analysis and generative tasks.
This transfer of NLP methods to symbolic music data has become more and more prevalent in the Music Information Retrieval (MIR) community, especially with the breakthrough of Transformer models.

\textit{Natural Language Processing} (NLP) is a field at the crossroads between linguistics and computer science that focuses on the interaction between computers and human language. 
Its main purpose is to understand, interpret, and generate human language taking into account its characteristics, such as syntactic or semantic properties.
Through various techniques, in particular, by training deep learning models, multiple tasks are tackled from text analysis such as sentiment analysis, part-of-speech tagging, text similarity or language identification to generative tasks such as summarization, question answering, chatbot conversation, or machine translation.

The field of \textit{Music Information Retrieval} (MIR) combines aspects of musicology and computer science to develop techniques to analyze music, retrieve music-related data, or generate music. 
While audio files capture music as sound, seen at a low level and described by objects such as waveforms or spectrograms, %
\textit{symbolic music} depicts music as abstract notations operating on concepts such as notes, chords, intervals, etc. that compose musical scores.
Although requiring more sophisticated notation systems,
symbolic music representations allow for the study of music at a higher level, such as analysis of harmony, form, or texture.
In practice, symbolic music remains prevalent in digital music production mainly relying on the MIDI format, which stands as a ubiquitous standard within digital audio workstations (DAWs). %
This survey will only consider music viewed as \textit{symbolic} representations.

\subsection{Music and natural language: similarities and specificities}

Beyond computer science studies, parallels between music and natural language are often drawn, as music is often considered as a linguistic system~\cite{jackendoff2009parallels}, sharing communication purposes as well as structural similarities with natural language.
These parallels are also found in terms of tasks studied in the NLP and MIR fields.

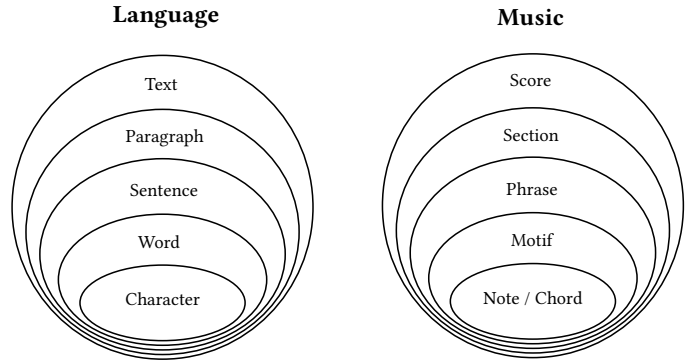
\begin{figure}[ht]
    \centering
    \resizebox{.6\linewidth}{!}{\tikzset{every picture/.style={line width=0.75pt}} %

\begin{tikzpicture}[x=0.75pt,y=0.75pt,yscale=-1,xscale=1]

\draw   (292,214) .. controls (292,200.47) and (315.95,189.5) .. (345.5,189.5) .. controls (375.05,189.5) and (399,200.47) .. (399,214) .. controls (399,227.53) and (375.05,238.5) .. (345.5,238.5) .. controls (315.95,238.5) and (292,227.53) .. (292,214) -- cycle ;
\draw   (278,199) .. controls (278,175.53) and (308.22,156.5) .. (345.5,156.5) .. controls (382.78,156.5) and (413,175.53) .. (413,199) .. controls (413,222.47) and (382.78,241.5) .. (345.5,241.5) .. controls (308.22,241.5) and (278,222.47) .. (278,199) -- cycle ;
\draw   (266,182.5) .. controls (266,148.26) and (301.59,120.5) .. (345.5,120.5) .. controls (389.41,120.5) and (425,148.26) .. (425,182.5) .. controls (425,216.74) and (389.41,244.5) .. (345.5,244.5) .. controls (301.59,244.5) and (266,216.74) .. (266,182.5) -- cycle ;
\draw   (257,168) .. controls (257,124.09) and (296.62,88.5) .. (345.5,88.5) .. controls (394.38,88.5) and (434,124.09) .. (434,168) .. controls (434,211.91) and (394.38,247.5) .. (345.5,247.5) .. controls (296.62,247.5) and (257,211.91) .. (257,168) -- cycle ;
\draw   (248,152) .. controls (248,97.6) and (291.65,53.5) .. (345.5,53.5) .. controls (399.35,53.5) and (443,97.6) .. (443,152) .. controls (443,206.4) and (399.35,250.5) .. (345.5,250.5) .. controls (291.65,250.5) and (248,206.4) .. (248,152) -- cycle ;
\draw   (52,215) .. controls (52,201.47) and (75.95,190.5) .. (105.5,190.5) .. controls (135.05,190.5) and (159,201.47) .. (159,215) .. controls (159,228.53) and (135.05,239.5) .. (105.5,239.5) .. controls (75.95,239.5) and (52,228.53) .. (52,215) -- cycle ;
\draw   (38,200) .. controls (38,176.53) and (68.22,157.5) .. (105.5,157.5) .. controls (142.78,157.5) and (173,176.53) .. (173,200) .. controls (173,223.47) and (142.78,242.5) .. (105.5,242.5) .. controls (68.22,242.5) and (38,223.47) .. (38,200) -- cycle ;
\draw   (26,183.5) .. controls (26,149.26) and (61.59,121.5) .. (105.5,121.5) .. controls (149.41,121.5) and (185,149.26) .. (185,183.5) .. controls (185,217.74) and (149.41,245.5) .. (105.5,245.5) .. controls (61.59,245.5) and (26,217.74) .. (26,183.5) -- cycle ;
\draw   (17,169) .. controls (17,125.09) and (56.62,89.5) .. (105.5,89.5) .. controls (154.38,89.5) and (194,125.09) .. (194,169) .. controls (194,212.91) and (154.38,248.5) .. (105.5,248.5) .. controls (56.62,248.5) and (17,212.91) .. (17,169) -- cycle ;
\draw   (8,153) .. controls (8,98.6) and (51.65,54.5) .. (105.5,54.5) .. controls (159.35,54.5) and (203,98.6) .. (203,153) .. controls (203,207.4) and (159.35,251.5) .. (105.5,251.5) .. controls (51.65,251.5) and (8,207.4) .. (8,153) -- cycle ;

\draw (107.86,29.51) node  [font=\LARGE] [align=left] {\begin{minipage}[lt]{78.88pt}\setlength\topsep{0pt}
\begin{center}
\textbf{Language}
\end{center}

\end{minipage}};
\draw (344.14,29.5) node  [font=\LARGE] [align=left] {\begin{minipage}[lt]{89.57pt}\setlength\topsep{0pt}
\begin{center}
\textbf{Music}
\end{center}

\end{minipage}};
\draw (106.93,107.84) node   [align=left] {\begin{minipage}[lt]{81.7pt}\setlength\topsep{0pt}
\begin{center}
Paragraph
\end{center}

\end{minipage}};
\draw (106.54,142.01) node   [align=left] {\begin{minipage}[lt]{77.47pt}\setlength\topsep{0pt}
\begin{center}
Sentence
\end{center}

\end{minipage}};
\draw (103.51,175.55) node   [align=left] {\begin{minipage}[lt]{69.35pt}\setlength\topsep{0pt}
\begin{center}
Word
\end{center}

\end{minipage}};
\draw (105.5,212) node   [align=left] {\begin{minipage}[lt]{61.2pt}\setlength\topsep{0pt}
\begin{center}
Character
\end{center}

\end{minipage}};
\draw (344.43,105.84) node   [align=left] {\begin{minipage}[lt]{81.7pt}\setlength\topsep{0pt}
\begin{center}
Section
\end{center}

\end{minipage}};
\draw (345.04,141.01) node   [align=left] {\begin{minipage}[lt]{77.47pt}\setlength\topsep{0pt}
\begin{center}
Phrase
\end{center}

\end{minipage}};
\draw (344.76,174.09) node   [align=left] {\begin{minipage}[lt]{73.09pt}\setlength\topsep{0pt}
\begin{center}
Motif
\end{center}

\end{minipage}};
\draw (345.5,212) node   [align=left] {\begin{minipage}[lt]{61.2pt}\setlength\topsep{0pt}
\begin{center}
Note / Chord
\end{center}

\end{minipage}};
\draw (344.14,70.16) node   [align=left] {\begin{minipage}[lt]{89.57pt}\setlength\topsep{0pt}
\begin{center}
Score
\end{center}

\end{minipage}};
\draw (104.85,73.18) node   [align=left] {\begin{minipage}[lt]{85.2pt}\setlength\topsep{0pt}
\begin{center}
Text
\end{center}

\end{minipage}};

\end{tikzpicture}}
    \caption{
        An \textit{oversimplified} example of segmentation levels in text and symbolic music.
        Such segmentations can, however, include more or less fine-grained levels and their delimitations can be ambiguous (Section~\ref{sec:discussion}).
    }
    \Description{Multiple levels of segmentation in symbolic music and text.}
    \label{fig:music_text_segmentations}
\end{figure}

\sepsubparagraph
\subparagraph{Music seen as a linguistic system} $\cdot$
Text representations and symbolic music representations are both semiotic systems~\cite{chomsky1979human} based on arrangements of symbols.
Text is built on characters and written music can be retranscribed with a variety of symbols derived from various notation systems such as standard notation, numbered notation or tablatures.
Both can be represented as sequences of elements which can be segmented or grouped at different levels. 
Text can be segmented into characters, syntactic phrases and sentences, while music 
can typically be segmented into temporal units such as notes, motifs, musical phrases, or sections~\cite{lerdahl2013musical} as represented in Figure~\ref{fig:music_text_segmentations}.

Inspired by higher-level concepts in natural language such as grammar or syntax, 
multiple models of musical syntaxes have been proposed~\cite{bernstein1976unanswered,bod2002unified}.
Such musical grammars rely on intrinsic musical concepts such as tension and relaxation~\cite{lerdahl1996generative} or harmony~\cite{rohrmeier2011towards}.
These grammatical or syntactic rules lead to expectancy in both language and music~\cite{jackendoff2009parallels,pearce2018statistical}, inducing similar cognitive reactions for the interlocutor or the listener when they are being transgressed in both language~\cite{pulvermuller2007grammar} and music~\cite{besson2001comparison}. %
Beyond its formal description, %
both are specific to human species and are learned through imitation. 
Both can also be perceived as elements unfolding in time~\cite{zbikowski2009music} and can be deployed under two modalities: an annotated form (text, sheet music) and an auditory form (speech, musical performance)~\cite{fornas1997text}.

However, major distinctions and particularities still persist between music and text (Section~\ref{sec:discussion}), including polyphony (simultaneous musical events), rhythm (rigorous musical time grid) and the multimodal aspect of notes being characterized by multiple musical features (pitch, dynamics, etc.).

\pgfplotsset{%
  every axis legend/.append style ={%
    anchor = north west,%
    at = {(0.02,0.97)}%
  },
  /pgf/number format/.cd,
  use comma,
  1000 sep = {\,},
  min exponent for 1000 sep = 4,
}

\definecolor{tab-1}{HTML}{1f77b4} %
\definecolor{tab-2}{HTML}{ff7f0e} %
\definecolor{tab-3}{HTML}{2ca02c} %
\definecolor{tab-4}{HTML}{d62728} %
\definecolor{tab-5}{HTML}{9467bd} %
\definecolor{tab-6}{HTML}{8c564b} %
\definecolor{tab-7}{HTML}{e377c2} %
\definecolor{tab-8}{HTML}{7f7f7f} %
\definecolor{tab-9}{HTML}{bcbd22} %
\definecolor{tab-10}{HTML}{17becf} %

\begin{figure}[t]
   \centering

  \resizebox{.47\linewidth}{!}{
  \begin{tikzpicture}
    \begin{axis}[
            xtick={2000,2002,2004,2006,2008,2010,2012,2014,2016,2018,2020,2022},
            ylabel={\# publications including this word},
            no markers,
            cycle list name=color list,
            width=.9\columnwidth,height=7cm,
            xmin=1999, xmax=2024,
            ymin=-1, ymax=17,
            legend style={
              nodes={scale=1, transform shape},
              font=\ttfamily,
              legend cell align={left}
            }, 
            x tick label style={
              rotate=0,
              anchor=north,
              font=\large
            }
        ]

        \addplot[tab-1] table[x=published,y=recurrent,col sep=comma]{data/ismir-count.csv};
        \addplot[tab-2] table[x=published,y=lstm,col sep=comma]{data/ismir-count.csv};
        \addplot[tab-3] table[x=published,y=attention,col sep=comma]{data/ismir-count.csv};
        \addplot[tab-4] table[x=published,y=transformer,col sep=comma]{data/ismir-count.csv};
        \addplot[tab-5] table[x=published,y=language,col sep=comma]{data/ismir-count.csv};
        \addplot[tab-6] table[x=published,y=tokens,col sep=comma]{data/ismir-count.csv};
        \addplot[tab-7] table[x=published,y=naturallanguage,col sep=comma]{data/ismir-count.csv};
        \addplot[tab-8] table[x=published,y=diffusion,col sep=comma]{data/ismir-count.csv};
        \addplot[gray, opacity=0.5, dashed] coordinates {(2017, -20) (2017, 100)};

        \legend{
          recurrent,
          lstm,
          attention,
          transformer,
          language,
          tokens,
          natural language,
          diffusion
        }
        \node[rotate=90,gray,above,opacity=0.5] at (axis cs:2017,14) {\small Transformer};
    \end{axis}
  \end{tikzpicture}
  }
  \resizebox{.47\linewidth}{!}{
  \begin{tikzpicture}
    \begin{axis}[
            xtick={2000,2002,2004,2006,2008,2010,2012,2014,2016,2018,2020,2022},
            ylabel={\# publications returned by the arXiv API},
            no markers,
            cycle list name=color list,
            width=.9\columnwidth,height=7cm,
            xmin=1999, xmax=2024,
            ymin=-1, ymax=115,
            legend style={
              nodes={scale=1, transform shape},
              font=\ttfamily,
              legend cell align={left}
            }, 
            x tick label style={
              rotate=0,
              anchor=north,
              font=\large
            }
        ]

        \addplot[tab-1] table[x=published,y=recurrent,col sep=comma]{data/arxiv-count.csv};
        \addplot[tab-2] table[x=published,y=lstm,col sep=comma]{data/arxiv-count.csv};
        \addplot[tab-3] table[x=published,y=attention,col sep=comma]{data/arxiv-count.csv};
        \addplot[tab-4] table[x=published,y=transformer,col sep=comma]{data/arxiv-count.csv};
        \addplot[tab-5] table[x=published,y=text,col sep=comma]{data/arxiv-count.csv};
        \addplot[tab-6] table[x=published,y=tokens,col sep=comma]{data/arxiv-count.csv};
        \addplot[tab-7] table[x=published,y=naturallanguage,col sep=comma]{data/arxiv-count.csv};
        \addplot[tab-8] table[x=published,y=diffusion,col sep=comma]{data/arxiv-count.csv};
        \addplot[gray, opacity=0.5, dashed] coordinates {(2017, -20) (2017, 200)};

        \legend{
          recurrent,
          lstm,
          attention,
          transformer,
          text,
          tokens,
          natural language,
          diffusion
        }
        \node[rotate=90,gray,above,opacity=0.5] at (axis cs:2017,90) {\small Transformer};
    \end{axis}
  \end{tikzpicture}
  }

   \caption{
        Evolution of the number of articles containing NLP-related words.
        (Left) Number of ISMIR papers containing NLP-related words 
        in their abstracts from 2000 to 2023.
        (Right) Number of arXiv preprints returned by the API query ``music AND <term>''.
    }
  \Description{Evolution of the number of articles containing NLP-related words.}
 \label{fig:ismir_nlp_words}
 \end{figure}
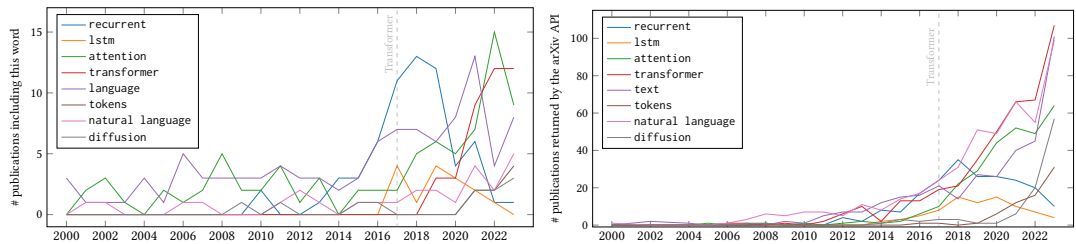

\sepsubparagraph
\subparagraph{Symbolic MIR and NLP tasks} $\cdot$
Beyond these parallels between text and symbolic music representations, the Natural Language Processing and Music Information Retrieval research fields are also related by similar tasks they address.
On the one hand, commong tasks on \textit{labeled} data in classification of whole textual document or music piece are common tasks, such as music composer classification~\cite{pollastri2001classification} and text authorship attribution~\cite{stamatatos2009survey}, folk song origin classification~\cite{hillewaere2009global} and language detection~\cite{jauhiainen2019automatic}, music genre~\cite{correa2016classification} and text style~\cite{kessler1997automatic} classification, or music emotion~\cite{hung2021emopia} and sentiment~\cite{wankhade2022survey} classification.
At a lower level, such labels can also describe textual or musical segments which naturally leads to a variety of segmentation tasks in both domains, including musical phrase retrieval~\cite{guan2018melodic} or musical form analysis~\cite{zhao2023forms} in MIR and discourse parsing~\cite{lin2019unified} or phrase segmentation~\cite{huang2010classical} in NLP.

On the other hand, tasks can rely on \textit{unlabeled} music and text datasets.
Apart from clustering tasks in text~\cite{xu2015short} and music~\cite{cilibrasi2004clustering}, these datasets are usually used to train generative systems following a self-supervised way (\ie predicting parts of the input itself, by learning representations and patterns without external annotations).
These models can be trained on tasks such as symbolic music infilling~\cite{guo2022musiac} and text infilling~\cite{donahue2020enabling}, %
or music priming~\cite{huang2018music} and text continuation~\cite{radford2019language}.
At the scale of a piece or a document, style transfer is performed in both MIR, through musical genres~\cite{wu2023musemorphose}, and NLP, through language high-level elements such as formality or toxicity~\cite{jin2022deep}.
More recently, text-conditioned generation has become more and more popular for the general public, including chatbot dialog\footnote{\url{https://chat.openai.com}} in NLP, and text-conditioned music generation~\cite{lu2023musecoco}.

These two fields also include numerous tasks that are inherent to one field, as depicted in Figure~\ref{fig:survey_overview}. 
These tasks specific to each field also reflect the inherent differences between these two types of data, including semantics in text which is key in an entailment task, or polyphony in music which is at the heart of harmonization and accompaniment generation tasks. 

\subsection{Applying NLP methods in symbolic MIR}

\begin{figure}[htbp]
    \centering
    \resizebox{.75\linewidth}{!}{\tikzset{every picture/.style={line width=0.75pt}} %

\begin{tikzpicture}[x=0.75pt,y=0.75pt,yscale=-1,xscale=1]

\draw  [fill={rgb, 255:red, 220; green, 220; blue, 220 }  ,fill opacity=1 ] (20.54,455.97) .. controls (20.54,450.15) and (25.25,445.43) .. (31.07,445.43) -- (310.94,445.43) .. controls (316.76,445.43) and (321.48,450.15) .. (321.48,455.97) -- (321.48,543.56) .. controls (321.48,549.38) and (316.76,554.1) .. (310.94,554.1) -- (31.07,554.1) .. controls (25.25,554.1) and (20.54,549.38) .. (20.54,543.56) -- cycle ;
\draw   (4,229) -- (336.22,229) -- (336.22,571.83) -- (4,571.83) -- cycle ;
\draw   (364.78,229) -- (697,229) -- (697,571.83) -- (364.78,571.83) -- cycle ;
\draw  [fill={rgb, 255:red, 201; green, 229; blue, 255 }  ,fill opacity=1 ] (391.25,395.34) .. controls (391.25,386.17) and (398.68,378.74) .. (407.85,378.74) -- (594.69,378.74) .. controls (603.86,378.74) and (611.29,386.17) .. (611.29,395.34) -- (611.29,445.14) .. controls (611.29,454.31) and (603.86,461.74) .. (594.69,461.74) -- (407.85,461.74) .. controls (398.68,461.74) and (391.25,454.31) .. (391.25,445.14) -- cycle ;
\draw  [fill={rgb, 255:red, 194; green, 224; blue, 191 }  ,fill opacity=1 ] (391,486.18) .. controls (391,477.02) and (398.43,469.58) .. (407.6,469.58) -- (594.44,469.58) .. controls (603.61,469.58) and (611.04,477.02) .. (611.04,486.18) -- (611.04,535.98) .. controls (611.04,545.15) and (603.61,552.58) .. (594.44,552.58) -- (407.6,552.58) .. controls (398.43,552.58) and (391,545.15) .. (391,535.98) -- cycle ;
\draw  [fill={rgb, 255:red, 255; green, 237; blue, 201 }  ,fill opacity=1 ] (657.14,381.47) .. controls (661.55,381.47) and (665.14,385.05) .. (665.14,389.47) -- (665.14,542.88) .. controls (665.14,547.3) and (661.55,550.88) .. (657.14,550.88) -- (633.14,550.88) .. controls (628.72,550.88) and (625.14,547.3) .. (625.14,542.88) -- (625.14,389.47) .. controls (625.14,385.05) and (628.72,381.47) .. (633.14,381.47) -- cycle ;
\draw  [fill={rgb, 255:red, 220; green, 220; blue, 220 }  ,fill opacity=1 ] (20.54,280.56) .. controls (20.54,271.67) and (27.74,264.47) .. (36.63,264.47) -- (305.39,264.47) .. controls (314.28,264.47) and (321.48,271.67) .. (321.48,280.56) -- (321.48,414.33) .. controls (321.48,423.22) and (314.28,430.42) .. (305.39,430.42) -- (36.63,430.42) .. controls (27.74,430.42) and (20.54,423.22) .. (20.54,414.33) -- cycle ;
\draw  [fill={rgb, 255:red, 207; green, 228; blue, 209 }  ,fill opacity=1 ] (29.56,310.25) .. controls (29.56,304.3) and (34.39,299.48) .. (40.34,299.48) -- (157.72,299.48) .. controls (163.68,299.48) and (168.5,304.3) .. (168.5,310.25) -- (168.5,402.37) .. controls (168.5,408.32) and (163.68,413.15) .. (157.72,413.15) -- (40.34,413.15) .. controls (34.39,413.15) and (29.56,408.32) .. (29.56,402.37) -- cycle ;
\draw  [draw opacity=0][fill={rgb, 255:red, 255; green, 196; blue, 196 }  ,fill opacity=0.5 ] (8,121.83) .. controls (8,79.95) and (121.05,46) .. (260.5,46) .. controls (399.95,46) and (513,79.95) .. (513,121.83) .. controls (513,163.71) and (399.95,197.66) .. (260.5,197.66) .. controls (121.05,197.66) and (8,163.71) .. (8,121.83) -- cycle ;
\draw  [draw opacity=0][fill={rgb, 255:red, 210; green, 219; blue, 255 }  ,fill opacity=0.5 ] (188,121.83) .. controls (188,79.95) and (301.05,46) .. (440.5,46) .. controls (579.95,46) and (693,79.95) .. (693,121.83) .. controls (693,163.71) and (579.95,197.66) .. (440.5,197.66) .. controls (301.05,197.66) and (188,163.71) .. (188,121.83) -- cycle ;
\draw   (8,121.83) .. controls (8,79.95) and (121.05,46) .. (260.5,46) .. controls (399.95,46) and (513,79.95) .. (513,121.83) .. controls (513,163.71) and (399.95,197.66) .. (260.5,197.66) .. controls (121.05,197.66) and (8,163.71) .. (8,121.83) -- cycle ;
\draw   (188,121.83) .. controls (188,79.95) and (301.05,46) .. (440.5,46) .. controls (579.95,46) and (693,79.95) .. (693,121.83) .. controls (693,163.71) and (579.95,197.66) .. (440.5,197.66) .. controls (301.05,197.66) and (188,163.71) .. (188,121.83) -- cycle ;
\draw  [fill={rgb, 255:red, 207; green, 228; blue, 209 }  ,fill opacity=1 ] (173.56,310.25) .. controls (173.56,304.3) and (178.39,299.48) .. (184.34,299.48) -- (301.72,299.48) .. controls (307.68,299.48) and (312.5,304.3) .. (312.5,310.25) -- (312.5,402.37) .. controls (312.5,408.32) and (307.68,413.15) .. (301.72,413.15) -- (184.34,413.15) .. controls (178.39,413.15) and (173.56,408.32) .. (173.56,402.37) -- cycle ;
\draw  [fill={rgb, 255:red, 207; green, 228; blue, 209 }  ,fill opacity=1 ] (29.56,485.87) .. controls (29.56,482.87) and (31.99,480.44) .. (34.99,480.44) -- (163.07,480.44) .. controls (166.07,480.44) and (168.5,482.87) .. (168.5,485.87) -- (168.5,532.3) .. controls (168.5,535.3) and (166.07,537.73) .. (163.07,537.73) -- (34.99,537.73) .. controls (31.99,537.73) and (29.56,535.3) .. (29.56,532.3) -- cycle ;
\draw  [fill={rgb, 255:red, 207; green, 228; blue, 209 }  ,fill opacity=1 ] (173.56,485.87) .. controls (173.56,482.87) and (175.99,480.44) .. (178.99,480.44) -- (307.07,480.44) .. controls (310.07,480.44) and (312.5,482.87) .. (312.5,485.87) -- (312.5,532.3) .. controls (312.5,535.3) and (310.07,537.73) .. (307.07,537.73) -- (178.99,537.73) .. controls (175.99,537.73) and (173.56,535.3) .. (173.56,532.3) -- cycle ;
\draw  [fill={rgb, 255:red, 255; green, 201; blue, 201 }  ,fill opacity=1 ] (383.24,276.97) .. controls (383.24,272.3) and (387.03,268.51) .. (391.71,268.51) -- (663.87,268.51) .. controls (668.54,268.51) and (672.33,272.3) .. (672.33,276.97) -- (672.33,302.37) .. controls (672.33,307.05) and (668.54,310.84) .. (663.87,310.84) -- (391.71,310.84) .. controls (387.03,310.84) and (383.24,307.05) .. (383.24,302.37) -- cycle ;
\draw  [fill={rgb, 255:red, 241; green, 143; blue, 143 }  ,fill opacity=1 ] (383.91,331.54) .. controls (383.91,326.87) and (387.7,323.08) .. (392.38,323.08) -- (664.53,323.08) .. controls (669.21,323.08) and (673,326.87) .. (673,331.54) -- (673,356.94) .. controls (673,361.62) and (669.21,365.41) .. (664.53,365.41) -- (392.38,365.41) .. controls (387.7,365.41) and (383.91,361.62) .. (383.91,356.94) -- cycle ;

\draw (263,7) node [anchor=north west][inner sep=0.75pt]  [font=\normalsize] [align=left] {\begin{minipage}[lt]{163.26pt}\setlength\topsep{0pt}
\begin{center}
\textbf{{\LARGE Tasks \& Evaluation}}
\end{center}

\end{minipage}};
\draw (83.88,235.49) node [anchor=north west][inner sep=0.75pt]  [font=\normalsize] [align=left] {\begin{minipage}[lt]{141.01pt}\setlength\topsep{0pt}
\begin{center}
\textbf{{\LARGE Representations}}
\end{center}

\end{minipage}};
\draw (508.66,234.55) node [anchor=north west][inner sep=0.75pt]  [font=\normalsize] [align=left] {\textbf{{\LARGE Models}}};
\draw (163,64) node [anchor=north west][inner sep=0.75pt]  [font=\Large] [align=left] {\textit{MIR}};
\draw (519,66) node [anchor=north west][inner sep=0.75pt]  [font=\Large] [align=left] {\textit{NLP}};
\draw (233,61.05) node [anchor=north west][inner sep=0.75pt]  [font=\large] [align=left] {\begin{minipage}[lt]{176.92pt}\setlength\topsep{0pt}
\begin{center}
Free generation\\Priming\\Infilling\\Style transfer\\Composer/author detection\\Style/genre classification\\Emotion/sentiment classification\\
\end{center}

\end{minipage}};
\draw (40,94.3) node [anchor=north west][inner sep=0.75pt]  [font=\footnotesize] [align=left] {Harmonic analysis\\Harmonization\\Accompaniment generation\\Text-to-MIDI\\...};
\draw (495.42,76.16) node [anchor=north west][inner sep=0.75pt]  [font=\footnotesize] [align=left] {\begin{minipage}[lt]{103.21pt}\setlength\topsep{0pt}
\begin{flushright}
Summarization\\Translation\\Entailment\\Part-of-speech tagging\\Named entity recognition\\Dependency parsing\\Word sense disambiguation\\...
\end{flushright}

\end{minipage}};
\draw (136.64,456.75) node [anchor=north west][inner sep=0.75pt]  [font=\large] [align=left] {Vectorization};
\draw (394.06,283.82) node [anchor=north west][inner sep=0.75pt]  [font=\large] [align=left] {Recurrent models};
\draw (405.5,402.7) node [anchor=north west][inner sep=0.75pt]  [font=\large] [align=left] {Training paradigm: \\\textit{- End-to-end / Pre-training}};
\draw (638.3,542.91) node [anchor=north west][inner sep=0.75pt]  [font=\large,rotate=-270] [align=left] {\begin{minipage}[lt]{114.98pt}\setlength\topsep{0pt}
\begin{center}
Mechanisms for MIR
\end{center}

\end{minipage}};
\draw (89.14,273.97) node [anchor=north west][inner sep=0.75pt]  [font=\large] [align=left] {\begin{minipage}[lt]{117.74pt}\setlength\topsep{0pt}
\begin{center}
Tokenization strategy
\end{center}

\end{minipage}};
\draw (46.36,313.76) node [anchor=north west][inner sep=0.75pt]  [font=\large] [align=left] {\begin{minipage}[lt]{74.87pt}\setlength\topsep{0pt}
\begin{center}
\textit{Event-based }\\\textit{tokenization}
\end{center}

\end{minipage}};
\draw (173.47,340.2) node [anchor=north west][inner sep=0.75pt]  [font=\large] [align=left] {\begin{minipage}[lt]{99.23pt}\setlength\topsep{0pt}
\begin{center}
\textit{Time-slice-based }\\\textit{tokenization}
\end{center}

\end{minipage}};
\draw (27.86,500.4) node [anchor=north west][inner sep=0.75pt]  [font=\large] [align=left] {\begin{minipage}[lt]{103.45pt}\setlength\topsep{0pt}
\begin{center}
\textit{Static embeddings}
\end{center}

\end{minipage}};
\draw (154.16,501.8) node [anchor=north west][inner sep=0.75pt]  [font=\large] [align=left] {\begin{minipage}[lt]{131.38pt}\setlength\topsep{0pt}
\begin{center}
\textit{Contextual embeddings}
\end{center}

\end{minipage}};
\draw (34,357.27) node [anchor=north west][inner sep=0.75pt]  [font=\large] [align=left] {$\displaystyle \rightsquigarrow $ \ Elementary tokens\\$\displaystyle \rightsquigarrow $ \ Composite tokens};
\draw (405.1,486.11) node [anchor=north west][inner sep=0.75pt]  [font=\large] [align=left] {Model architecture: \\\textit{- Encoder / Decoder}\\\textit{- Multimodal models}};
\draw (393.39,337.17) node [anchor=north west][inner sep=0.75pt]  [font=\large] [align=left] {Attention-based models};

\end{tikzpicture}}
    \caption{
        Overview of the survey, organized around two axes:
        similarities and specificities of NLP and MIR tasks motivating
        \textit{representations} of symbolic music inspired by NLP and
        \textit{models} adapted from NLP for symbolic music.
    }
    \Description{Overview of the survey}
    \label{fig:survey_overview}
\end{figure}
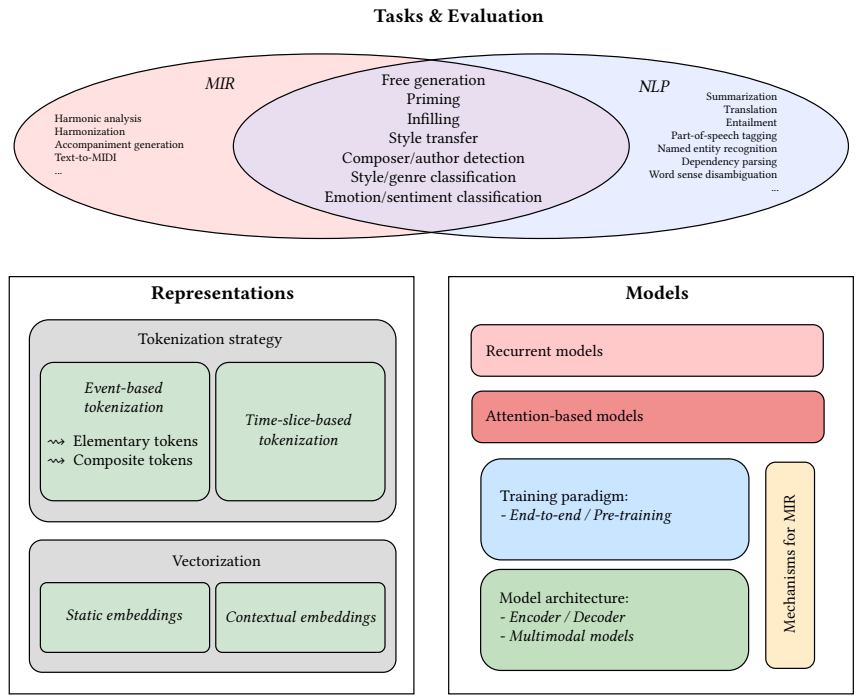

In the field of symbolic Music Information Retrieval, multiple surveys have been published with a clear focus on music generation.
Two main classes of surveys seem to emerge: ones presenting systems through their technical aspects, and ones %
organizing them based on their musical purpose or task.
Firstly, various surveys are driven by the system's technical aspects~\cite{fernandez2013ai}.
More recent surveys now specifically focus on deep learning methods~\cite{briot2020deep}.
These surveys organize deep learning generative models following multiple axes such as model architecture~\cite{wang2023review}, types of generation conditions~\cite{zhu2023survey}, and emotion-driven generation~\cite{dash2023aibased}.
Instead, several overview articles focus on musical tasks~\cite{herremans2017functional} and categorize them based on the nature of the generated content~\cite{liu2016computational} or by the conditions imposed for generation tasks~\cite{ji2023survey}.

An observation drawn from these surveys indicates that the MIR community is closely following new advances in NLP by adapting their tools for music purposes. 
MIR studies are used to adapting techniques from other fields, such as image processing~\cite{huang2017counterpoint}, %
resulting in this current trend regarding NLP methods for symbolic MIR.
Figure~\ref{fig:ismir_nlp_words} describes the number of publications from the ISMIR conference %
that include NLP-related terms in their abstract as well as
music/NLP-related arXiv preprints.
In particular, from 2017 with the introduction of Transformers, 
references to NLP techniques or models have increased drastically so that most of the state-of-the-art models in symbolic music tasks are now based on this model.
This phenomenon has encouraged dedicated initiatives in the MIR community,
such as the organization of the workshop NLP4MuSA (Workshop on NLP for Music and Spoken Audio)\footnote{\url{https://sites.google.com/view/nlp4musa}} held in 2020 and 2021.
In addition, more and more overviews of deep learning approaches for music generation, including NLP-based methods, are presented as tutorials at conferences such as ISMIR\footnote{\url{http://ismir2023program.ismir.net/tutorials.html\#T3}} or CMMR\footnote{\url{https://cmmr2023.gttm.jp/keynotes/\#Yang_abst}}.

The original approach introduced in this survey emphasizes the adaptation of Natural Language Processing methods for music generation and information retrieval within the domain of symbolic music.
These encompass tools and methods not only for symbolic music generation but also for existing analysis tasks.
From a more epistemological point of view, we hope that analyzing NLP approaches to process symbolic music representations brings an original and promising approach to reconsider the question of what music shares with natural language.

We present an overview of NLP methods adapted for symbolic MIR by proposing taxonomies of two technical aspects (Figure~\ref{fig:survey_overview}): \textit{representations} (Section~\ref{sec:representations}) and \textit{models} (Section~\ref{sec:models}).

\begin{itemize}[beginpenalty=10000]
    \item Choosing a \textbf{representation} refers to encoding content (text or symbolic music) into a format suitable for computational processing. Adapting NLP models to symbolic MIR involves mainly sequential representations.
    \item The \textbf{model} performs the task by processing a \textit{representation} of the input content. Such a model can be based on recurrent layers or attention heads, with specific architectures or training paradigms, and potentially implements mechanisms tailored for symbolic music data.
\end{itemize}

We then discuss the use of such NLP techniques for symbolic MIR by raising possible technical limitations when employing these methods and numerous differences between music and text. We finally outline future directions in which NLP methods can be implemented and adapted for symbolic music (Section~\ref{sec:discussion}).

New models or methods adapted from NLP to MIR are released extremely frequently: this survey includes such developments up until the end of 2023.
To facilitate continuous updates with these newly released tools, we maintain a collaborative repository accessible at: \url{https://github.com/dinhviettoanle/survey-music-nlp}.

\section{Representations of symbolic music as sequences}
\label{sec:representations}

Text data inherently follows a sequential structure composed of elements
spanning from individual characters to full sentences.
In contrast, %
representing musical content as a sequence of homogeneous elements is not as straightforward.
The multiplicity of information included in a single note (pitch, duration, position, etc.) and the common occurrences of simultaneous notes (polyphony, chords and melody, etc.) make the problem more complex than with text.
However, this sequential representation is necessary for the musical data to be subsequently processed by sequential models, which were initially designed to handle text data. 
This section presents various methods that have been developed to represent \textit{music as sequences of elements}.

\subsection{Tokenization strategies}

\begin{figure}[ht]
    \centering
    \resizebox{.7\linewidth}{!}{\tikzset{every picture/.style={line width=0.75pt}} %

\begin{tikzpicture}[x=0.75pt,y=0.75pt,yscale=-1,xscale=1]

\draw  [fill={rgb, 255:red, 244; green, 244; blue, 244 }  ,fill opacity=1 ]  (6,51.75) .. controls (6,48.99) and (8.24,46.75) .. (11,46.75) -- (199,46.75) .. controls (201.76,46.75) and (204,48.99) .. (204,51.75) -- (204,77.75) .. controls (204,80.51) and (201.76,82.75) .. (199,82.75) -- (11,82.75) .. controls (8.24,82.75) and (6,80.51) .. (6,77.75) -- cycle  ;
\draw (105,64.75) node  [font=\Large] [align=left] {\begin{minipage}[lt]{131.92pt}\setlength\topsep{0pt}
\begin{center}
\textbf{Tokenization strategies}
\end{center}

\end{minipage}};
\draw  [fill={rgb, 255:red, 217; green, 217; blue, 217 }  ,fill opacity=1 ]  (243,87.25) .. controls (243,84.49) and (245.24,82.25) .. (248,82.25) -- (436,82.25) .. controls (438.76,82.25) and (441,84.49) .. (441,87.25) -- (441,117.25) .. controls (441,120.01) and (438.76,122.25) .. (436,122.25) -- (248,122.25) .. controls (245.24,122.25) and (243,120.01) .. (243,117.25) -- cycle  ;
\draw (342,102.25) node  [font=\Large] [align=left] {\begin{minipage}[lt]{131.92pt}\setlength\topsep{0pt}
\begin{center}
\textbf{Event-based tokenization}
\end{center}

\end{minipage}};
\draw  [fill={rgb, 255:red, 190; green, 190; blue, 190 }  ,fill opacity=1 ]  (523,16.29) .. controls (523,13.53) and (525.24,11.29) .. (528,11.29) -- (716,11.29) .. controls (718.76,11.29) and (721,13.53) .. (721,16.29) -- (721,79.29) .. controls (721,82.05) and (718.76,84.29) .. (716,84.29) -- (528,84.29) .. controls (525.24,84.29) and (523,82.05) .. (523,79.29) -- cycle  ;
\draw (622,47.79) node   [align=left] {\begin{minipage}[lt]{131.92pt}\setlength\topsep{0pt}
\begin{center}
\end{center}

\end{minipage}};
\draw  [fill={rgb, 255:red, 190; green, 190; blue, 190 }  ,fill opacity=1 ]  (525,106) .. controls (525,103.24) and (527.24,101) .. (530,101) -- (718,101) .. controls (720.76,101) and (723,103.24) .. (723,106) -- (723,132) .. controls (723,134.76) and (720.76,137) .. (718,137) -- (530,137) .. controls (527.24,137) and (525,134.76) .. (525,132) -- cycle  ;
\draw (624,119) node  [font=\Large] [align=left] {\begin{minipage}[lt]{131.92pt}\setlength\topsep{0pt}
\begin{center}
\textbf{Composite tokens}
\end{center}

\end{minipage}};
\draw  [fill={rgb, 255:red, 163; green, 163; blue, 163 }  ,fill opacity=1 ]  (539.08,42.42) -- (616.08,42.42) -- (616.08,76.42) -- (539.08,76.42) -- cycle  ;
\draw (577.58,59.42) node  [font=\large] [align=left] {\begin{minipage}[lt]{49.43pt}\setlength\topsep{0pt}
\begin{center}
Alphabet
\end{center}

\end{minipage}};
\draw  [fill={rgb, 255:red, 163; green, 163; blue, 163 }  ,fill opacity=1 ]  (630.08,42.42) -- (707.08,42.42) -- (707.08,76.42) -- (630.08,76.42) -- cycle  ;
\draw (668.58,59.42) node  [font=\large] [align=left] {\begin{minipage}[lt]{49.43pt}\setlength\topsep{0pt}
\begin{center}
Grouping
\end{center}

\end{minipage}};
\draw  [fill={rgb, 255:red, 217; green, 217; blue, 217 }  ,fill opacity=1 ]  (243,10.25) .. controls (243,7.49) and (245.24,5.25) .. (248,5.25) -- (436,5.25) .. controls (438.76,5.25) and (441,7.49) .. (441,10.25) -- (441,40.25) .. controls (441,43.01) and (438.76,45.25) .. (436,45.25) -- (248,45.25) .. controls (245.24,45.25) and (243,43.01) .. (243,40.25) -- cycle  ;
\draw (342,25.25) node  [font=\Large] [align=left] {\begin{minipage}[lt]{131.92pt}\setlength\topsep{0pt}
\begin{center}
\textbf{Time-slice-based tokenization}
\end{center}

\end{minipage}};
\draw (533,18) node [anchor=north west][inner sep=0.75pt]  [font=\Large] [align=left] {\begin{minipage}[lt]{133.3pt}\setlength\topsep{0pt}
\begin{center}
\textbf{Elementary tokens}
\end{center}

\end{minipage}};
\draw    (204,80.41) -- (241.02,86.27) ;
\draw [shift={(243,86.59)}, rotate = 188.99] [color={rgb, 255:red, 0; green, 0; blue, 0 }  ][line width=0.75]    (10.93,-3.29) .. controls (6.95,-1.4) and (3.31,-0.3) .. (0,0) .. controls (3.31,0.3) and (6.95,1.4) .. (10.93,3.29)   ;
\draw    (441,83) -- (521.04,67.43) ;
\draw [shift={(523,67.05)}, rotate = 168.99] [color={rgb, 255:red, 0; green, 0; blue, 0 }  ][line width=0.75]    (10.93,-3.29) .. controls (6.95,-1.4) and (3.31,-0.3) .. (0,0) .. controls (3.31,0.3) and (6.95,1.4) .. (10.93,3.29)   ;
\draw    (441,108.13) -- (523,113) ;
\draw [shift={(525,113.12)}, rotate = 183.4] [color={rgb, 255:red, 0; green, 0; blue, 0 }  ][line width=0.75]    (10.93,-3.29) .. controls (6.95,-1.4) and (3.31,-0.3) .. (0,0) .. controls (3.31,0.3) and (6.95,1.4) .. (10.93,3.29)   ;
\draw    (204,48.25) -- (241.03,42.08) ;
\draw [shift={(243,41.75)}, rotate = 170.54] [color={rgb, 255:red, 0; green, 0; blue, 0 }  ][line width=0.75]    (10.93,-3.29) .. controls (6.95,-1.4) and (3.31,-0.3) .. (0,0) .. controls (3.31,0.3) and (6.95,1.4) .. (10.93,3.29)   ;

\end{tikzpicture}}
    \caption{
        Taxonomy of tokenizations for symbolic music.
        Tokens are either based on regular time-slices or events. 
        Among event-based tokenization strategies, tokens encode various features of these events: composite (or multidimensional) tokens encapsulate all these features in a single token, in contrast with elementary tokens where each musical feature is processed one after the other.
    }
    \Description{Tokenizations ontology}
    \label{fig:tokenization_overview}
\end{figure}
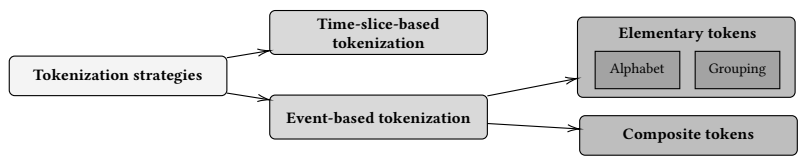

Tokenization refers to the process of representing complex content into a sequence of elements for computational processing. 
In NLP, tokenization is the task of segmenting a sequence of atomic elements - characters - by grouping them together into informative \textit{tokens}~\cite{mielke2021between}, such as subwords, words, or multiple-word expressions. The rise of NLP models in MIR has naturally encouraged the adoption of this term in music representations.
We propose a taxonomy of tokenization strategies in symbolic MIR represented in Figure~\ref{fig:tokenization_overview}.

We organize tokenization strategies within two classes: \textit{time-slice-based tokenization} and \textit{event-based tokenization}.
Time plays a special role in music since the time position of notes fundamentally contributes to the conveyed information.
Musical elements are intended to occur on an isochronic grid~\cite{jackendoff2009parallels} in which notes have rigorous annotated timings on sheet music\footnote{
Such exact timings can, however, be altered in a performance context where musicians have the freedom to distort this time grid leading to expressive effects such as \textit{rubato}, \textit{accelerando}, or \textit{ritardando}.
}.
Representing time properties of musical elements has led to multiple approaches~\cite[$\S$4.8]{briot2020deep} including representations based on regular time steps (Section~\ref{sec:time_slice}), or driven by events occurring through time (Section~\ref{sec:event_based}).

\subsubsection{Time-slice-based tokenization}
\label{sec:time_slice}
Dividing time at evenly spaced timings is a natural approach to representing music since musical elements are notated on scores at specific timings according to particular rhythms. The approaches described in the following section represent symbolic music as a sequence of fixed-time interval tokens.

DeepBach~\cite{hadjeres2017deepbach} is a model that aims to generate 4-part chorales, for which time is evenly divided at the level of 16th notes.
As the number of simultaneous notes is upper-bounded in 4-part chorales, a time step can be represented as a vector containing 4 pitches. %
In the same way, a concept of ``musical words'' defined by slices of three beats is proposed~\cite{herremans2017modeling,chuan2020context} to model musical context and semantic relationships in polyphonic music.
Beyond pitches, this time-slice representation can be used in the context of drum music~\cite{zhang2023drummers}.
More generally, these representations can be seen as specific cases of piano rolls.
This representation relies on matrices in which the horizontal axis represents time, and pitches are encoded along the vertical axis, with additional characteristics such as velocity as a third dimension. 
Piano rolls are usually portrayed as an alternative to sequential representations by using matrices.
However, a piano roll can be converted into a sequential format by considering it as a sequence of piano roll slices - \ie fixed-size multi-hot vectors containing pitches quantized at a specific duration.
These serialized piano rolls 
consider tokens which can represent
a small window of slices around a middle piano roll slice~\cite{chen2019harmony}, 
or a full musical bar~\cite{brunner2018midi}.

\subsubsection{Event-based tokenization}
\label{sec:event_based}
Unlike time-slice-based tokenization in which tokens are triggered at each time step, \textit{event-based tokenization strategies} involve tokens occurring when a specific event takes place (\eg a note being played, the start of a measure, etc.).
Most tokenization strategies have shifted towards this event-centric approach, helped by the increasing amount of available MIDI data.
The MIDI protocol (Musical Instrument Digital Interface) was first developed to handle communication between music software and hardware. 
The serial transmission of MIDI messages provides a natural way to encode music as sequences of events. %
The large adoption of this format in the music community has led to the availability of multiple datasets~\cite{ji2023survey} which are essential for training deep learning models. %

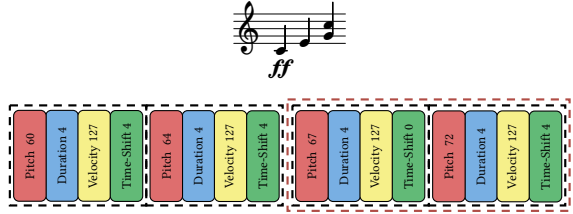
\begin{figure}[ht]
    \centering
    \includegraphics[width=1.5cm]{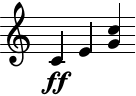}

    \vspace{.7em}
    
    \resizebox{.5\linewidth}{!}{\tikzset{every picture/.style={line width=0.75pt}} %

\begin{tikzpicture}[x=0.75pt,y=0.75pt,yscale=-1,xscale=1]

\draw  [dash pattern={on 5.63pt off 4.5pt}][line width=1.5]  (8,9.75) -- (134,9.75) -- (134,102.25) -- (8,102.25) -- cycle ;
\draw  [dash pattern={on 5.63pt off 4.5pt}][line width=1.5]  (134,9.75) -- (260,9.75) -- (260,102.25) -- (134,102.25) -- cycle ;
\draw  [dash pattern={on 5.63pt off 4.5pt}][line width=1.5]  (269,9.75) -- (395,9.75) -- (395,102.25) -- (269,102.25) -- cycle ;
\draw  [dash pattern={on 5.63pt off 4.5pt}][line width=1.5]  (395,9.75) -- (521,9.75) -- (521,102.25) -- (395,102.25) -- cycle ;
\draw  [color={rgb, 255:red, 176; green, 77; blue, 70 }  ,draw opacity=1 ][dash pattern={on 5.63pt off 4.5pt}][line width=1.5]  (264.5,4.25) -- (525,4.25) -- (525,107.25) -- (264.5,107.25) -- cycle ;

\draw  [fill={rgb, 255:red, 120; green, 175; blue, 227 }  ,fill opacity=1 ]  (41.22,19.34) .. controls (41.22,16.58) and (43.46,14.34) .. (46.22,14.34) -- (66.22,14.34) .. controls (68.98,14.34) and (71.22,16.58) .. (71.22,19.34) -- (71.22,93.34) .. controls (71.22,96.1) and (68.98,98.34) .. (66.22,98.34) -- (46.22,98.34) .. controls (43.46,98.34) and (41.22,96.1) .. (41.22,93.34) -- cycle  ;
\draw (56.22,56.34) node  [rotate=-270] [align=left] {\begin{minipage}[lt]{54.52pt}\setlength\topsep{0pt}
\begin{center}
Duration 4
\end{center}

\end{minipage}};
\draw  [fill={rgb, 255:red, 246; green, 240; blue, 138 }  ,fill opacity=1 ]  (70.94,19.34) .. controls (70.94,16.58) and (73.18,14.34) .. (75.94,14.34) -- (95.94,14.34) .. controls (98.71,14.34) and (100.94,16.58) .. (100.94,19.34) -- (100.94,93.34) .. controls (100.94,96.1) and (98.71,98.34) .. (95.94,98.34) -- (75.94,98.34) .. controls (73.18,98.34) and (70.94,96.1) .. (70.94,93.34) -- cycle  ;
\draw (85.94,56.34) node  [rotate=-270] [align=left] {\begin{minipage}[lt]{54.52pt}\setlength\topsep{0pt}
\begin{center}
Velocity 127
\end{center}

\end{minipage}};
\draw  [fill={rgb, 255:red, 222; green, 110; blue, 110 }  ,fill opacity=1 ]  (11.5,19.34) .. controls (11.5,16.58) and (13.74,14.34) .. (16.5,14.34) -- (36.5,14.34) .. controls (39.26,14.34) and (41.5,16.58) .. (41.5,19.34) -- (41.5,93.34) .. controls (41.5,96.1) and (39.26,98.34) .. (36.5,98.34) -- (16.5,98.34) .. controls (13.74,98.34) and (11.5,96.1) .. (11.5,93.34) -- cycle  ;
\draw (26.5,56.34) node  [rotate=-270] [align=left] {\begin{minipage}[lt]{54.52pt}\setlength\topsep{0pt}
\begin{center}
Pitch \ 60
\end{center}

\end{minipage}};
\draw  [fill={rgb, 255:red, 98; green, 186; blue, 119 }  ,fill opacity=1 ]  (100.72,19.5) .. controls (100.72,16.74) and (102.96,14.5) .. (105.72,14.5) -- (125.72,14.5) .. controls (128.48,14.5) and (130.72,16.74) .. (130.72,19.5) -- (130.72,93.5) .. controls (130.72,96.26) and (128.48,98.5) .. (125.72,98.5) -- (105.72,98.5) .. controls (102.96,98.5) and (100.72,96.26) .. (100.72,93.5) -- cycle  ;
\draw (115.72,56.5) node  [rotate=-270] [align=left] {\begin{minipage}[lt]{54.52pt}\setlength\topsep{0pt}
\begin{center}
Time-Shift 4
\end{center}

\end{minipage}};
\draw  [fill={rgb, 255:red, 120; green, 175; blue, 227 }  ,fill opacity=1 ]  (167.22,19.59) .. controls (167.22,16.83) and (169.46,14.59) .. (172.22,14.59) -- (192.22,14.59) .. controls (194.98,14.59) and (197.22,16.83) .. (197.22,19.59) -- (197.22,93.59) .. controls (197.22,96.35) and (194.98,98.59) .. (192.22,98.59) -- (172.22,98.59) .. controls (169.46,98.59) and (167.22,96.35) .. (167.22,93.59) -- cycle  ;
\draw (182.22,56.59) node  [rotate=-270] [align=left] {\begin{minipage}[lt]{54.52pt}\setlength\topsep{0pt}
\begin{center}
Duration 4
\end{center}

\end{minipage}};
\draw  [fill={rgb, 255:red, 246; green, 240; blue, 138 }  ,fill opacity=1 ]  (196.94,19.59) .. controls (196.94,16.83) and (199.18,14.59) .. (201.94,14.59) -- (221.94,14.59) .. controls (224.71,14.59) and (226.94,16.83) .. (226.94,19.59) -- (226.94,93.59) .. controls (226.94,96.35) and (224.71,98.59) .. (221.94,98.59) -- (201.94,98.59) .. controls (199.18,98.59) and (196.94,96.35) .. (196.94,93.59) -- cycle  ;
\draw (211.94,56.59) node  [rotate=-270] [align=left] {\begin{minipage}[lt]{54.52pt}\setlength\topsep{0pt}
\begin{center}
Velocity 127
\end{center}

\end{minipage}};
\draw  [fill={rgb, 255:red, 222; green, 110; blue, 110 }  ,fill opacity=1 ]  (137.5,19.59) .. controls (137.5,16.83) and (139.74,14.59) .. (142.5,14.59) -- (162.5,14.59) .. controls (165.26,14.59) and (167.5,16.83) .. (167.5,19.59) -- (167.5,93.59) .. controls (167.5,96.35) and (165.26,98.59) .. (162.5,98.59) -- (142.5,98.59) .. controls (139.74,98.59) and (137.5,96.35) .. (137.5,93.59) -- cycle  ;
\draw (152.5,56.59) node  [rotate=-270] [align=left] {\begin{minipage}[lt]{54.52pt}\setlength\topsep{0pt}
\begin{center}
Pitch \ 64
\end{center}

\end{minipage}};
\draw  [fill={rgb, 255:red, 98; green, 186; blue, 119 }  ,fill opacity=1 ]  (226.72,19.75) .. controls (226.72,16.99) and (228.96,14.75) .. (231.72,14.75) -- (251.72,14.75) .. controls (254.48,14.75) and (256.72,16.99) .. (256.72,19.75) -- (256.72,93.75) .. controls (256.72,96.51) and (254.48,98.75) .. (251.72,98.75) -- (231.72,98.75) .. controls (228.96,98.75) and (226.72,96.51) .. (226.72,93.75) -- cycle  ;
\draw (241.72,56.75) node  [rotate=-270] [align=left] {\begin{minipage}[lt]{54.52pt}\setlength\topsep{0pt}
\begin{center}
Time-Shift 4
\end{center}

\end{minipage}};
\draw  [fill={rgb, 255:red, 120; green, 175; blue, 227 }  ,fill opacity=1 ]  (301.72,19.59) .. controls (301.72,16.83) and (303.96,14.59) .. (306.72,14.59) -- (326.72,14.59) .. controls (329.48,14.59) and (331.72,16.83) .. (331.72,19.59) -- (331.72,93.59) .. controls (331.72,96.35) and (329.48,98.59) .. (326.72,98.59) -- (306.72,98.59) .. controls (303.96,98.59) and (301.72,96.35) .. (301.72,93.59) -- cycle  ;
\draw (316.72,56.59) node  [rotate=-270] [align=left] {\begin{minipage}[lt]{54.52pt}\setlength\topsep{0pt}
\begin{center}
Duration 4
\end{center}

\end{minipage}};
\draw  [fill={rgb, 255:red, 246; green, 240; blue, 138 }  ,fill opacity=1 ]  (331.44,19.59) .. controls (331.44,16.83) and (333.68,14.59) .. (336.44,14.59) -- (356.44,14.59) .. controls (359.21,14.59) and (361.44,16.83) .. (361.44,19.59) -- (361.44,93.59) .. controls (361.44,96.35) and (359.21,98.59) .. (356.44,98.59) -- (336.44,98.59) .. controls (333.68,98.59) and (331.44,96.35) .. (331.44,93.59) -- cycle  ;
\draw (346.44,56.59) node  [rotate=-270] [align=left] {\begin{minipage}[lt]{54.52pt}\setlength\topsep{0pt}
\begin{center}
Velocity 127
\end{center}

\end{minipage}};
\draw  [fill={rgb, 255:red, 222; green, 110; blue, 110 }  ,fill opacity=1 ]  (272,19.59) .. controls (272,16.83) and (274.24,14.59) .. (277,14.59) -- (297,14.59) .. controls (299.76,14.59) and (302,16.83) .. (302,19.59) -- (302,93.59) .. controls (302,96.35) and (299.76,98.59) .. (297,98.59) -- (277,98.59) .. controls (274.24,98.59) and (272,96.35) .. (272,93.59) -- cycle  ;
\draw (287,56.59) node  [rotate=-270] [align=left] {\begin{minipage}[lt]{54.52pt}\setlength\topsep{0pt}
\begin{center}
Pitch \ 67
\end{center}

\end{minipage}};
\draw  [fill={rgb, 255:red, 98; green, 186; blue, 119 }  ,fill opacity=1 ]  (361.22,19.75) .. controls (361.22,16.99) and (363.46,14.75) .. (366.22,14.75) -- (386.22,14.75) .. controls (388.98,14.75) and (391.22,16.99) .. (391.22,19.75) -- (391.22,93.75) .. controls (391.22,96.51) and (388.98,98.75) .. (386.22,98.75) -- (366.22,98.75) .. controls (363.46,98.75) and (361.22,96.51) .. (361.22,93.75) -- cycle  ;
\draw (376.22,56.75) node  [rotate=-270] [align=left] {\begin{minipage}[lt]{54.52pt}\setlength\topsep{0pt}
\begin{center}
Time-Shift 0
\end{center}

\end{minipage}};
\draw  [fill={rgb, 255:red, 120; green, 175; blue, 227 }  ,fill opacity=1 ]  (428.22,19.84) .. controls (428.22,17.08) and (430.46,14.84) .. (433.22,14.84) -- (453.22,14.84) .. controls (455.98,14.84) and (458.22,17.08) .. (458.22,19.84) -- (458.22,93.84) .. controls (458.22,96.6) and (455.98,98.84) .. (453.22,98.84) -- (433.22,98.84) .. controls (430.46,98.84) and (428.22,96.6) .. (428.22,93.84) -- cycle  ;
\draw (443.22,56.84) node  [rotate=-270] [align=left] {\begin{minipage}[lt]{54.52pt}\setlength\topsep{0pt}
\begin{center}
Duration 4
\end{center}

\end{minipage}};
\draw  [fill={rgb, 255:red, 246; green, 240; blue, 138 }  ,fill opacity=1 ]  (457.94,19.84) .. controls (457.94,17.08) and (460.18,14.84) .. (462.94,14.84) -- (482.94,14.84) .. controls (485.71,14.84) and (487.94,17.08) .. (487.94,19.84) -- (487.94,93.84) .. controls (487.94,96.6) and (485.71,98.84) .. (482.94,98.84) -- (462.94,98.84) .. controls (460.18,98.84) and (457.94,96.6) .. (457.94,93.84) -- cycle  ;
\draw (472.94,56.84) node  [rotate=-270] [align=left] {\begin{minipage}[lt]{54.52pt}\setlength\topsep{0pt}
\begin{center}
Velocity 127
\end{center}

\end{minipage}};
\draw  [fill={rgb, 255:red, 222; green, 110; blue, 110 }  ,fill opacity=1 ]  (398.5,19.84) .. controls (398.5,17.08) and (400.74,14.84) .. (403.5,14.84) -- (423.5,14.84) .. controls (426.26,14.84) and (428.5,17.08) .. (428.5,19.84) -- (428.5,93.84) .. controls (428.5,96.6) and (426.26,98.84) .. (423.5,98.84) -- (403.5,98.84) .. controls (400.74,98.84) and (398.5,96.6) .. (398.5,93.84) -- cycle  ;
\draw (413.5,56.84) node  [rotate=-270] [align=left] {\begin{minipage}[lt]{54.52pt}\setlength\topsep{0pt}
\begin{center}
Pitch \ 72
\end{center}

\end{minipage}};
\draw  [fill={rgb, 255:red, 98; green, 186; blue, 119 }  ,fill opacity=1 ]  (487.72,20) .. controls (487.72,17.24) and (489.96,15) .. (492.72,15) -- (512.72,15) .. controls (515.48,15) and (517.72,17.24) .. (517.72,20) -- (517.72,94) .. controls (517.72,96.76) and (515.48,99) .. (512.72,99) -- (492.72,99) .. controls (489.96,99) and (487.72,96.76) .. (487.72,94) -- cycle  ;
\draw (502.72,57) node  [rotate=-270] [align=left] {\begin{minipage}[lt]{54.52pt}\setlength\topsep{0pt}
\begin{center}
Time-Shift 4
\end{center}

\end{minipage}};

\end{tikzpicture}}
    \caption{
        Artificial sequentiality possibly introduced in a tokenization strategy. 
        By restraining the attributes of a note to pitch, duration, velocity, and time-shift, the sequentiality of the blocks (black dashed blocks) follows the temporality, but the order of the inner musical features is arbitrary.
        The sequentiality of these blocks can even be artificial for simultaneous events (red dashed block).
    }
    \Description{Artificial sequentiality in tokenization}
    \label{fig:artificial_sequentiality}
\end{figure}

However, MIDI messages are of different types: unlike characters in text, musical notes include multiple features, such as duration, pitch, or velocity.
Since these features characterize a single temporal event, representing such features sequentially may necessitate introducing an ``artificial'' sequentiality on top of the temporal sequentiality as illustrated in Figure~\ref{fig:artificial_sequentiality}. 
This sequentiality is even more artificial when representing simultaneous notes occurring at the same time.
In the MIR field, two main classes of event-based tokens stand out that we refer to as \textit{elementary tokens} (Table~\ref{tab:tokenizations_elementary}) and \textit{composite tokens} (Table~\ref{tab:tokenizations_composite}).
Sequences of elementary tokens explicitly integrate this artificial sequentiality where each token is a single musical feature.
This can possibly result in two adjacent tokens describing the same temporal event (\eg the pitch of a note followed by its duration).
On the contrary, sequences of composite tokens partly bypass this artificial sequentiality by considering tokens as objects aggregating all the musical features describing a temporal event in a unique ``super-token''.

\paragraph{Elementary tokens -- Music as sequence of individual features} %
The constitutive elements of a sequence composed of musical elementary tokens can be described at two levels (Table~\ref{tab:tokenizations_elementary}): the choices of an initial \textit{alphabet} of atomic elements encoding different musical features and a \textit{grouping} of these atomic elements into higher level elements, presumably more expressive.

\afterpage{
    \newcommand{\headTokenizationElementary}{
    {\textbf{Tokenization}} & 
    {\textbf{\makecell{Score-based /\\Perf.-based}}} & 
    {\textbf{\scriptsize\makecell[c]{Alphabet \itshape(Atomic elements)}}} & 
    {\textbf{Grouping}} & 
    {\textbf{\makecell{Vocab.\\size}}} &
    {\textbf{\makecell{Data}}}\\
    \headtoprule
}

{
    \scriptsize
    \renewcommand\cellalign{cl}
    \def\arraystretch{1.3}

    \begin{longtable}{ll>{\tiny}llll}
    \caption[]{
        Overview\footnotemark of event-based tokenization strategies based on \textit{elementary} tokens.
        The ``alphabet'' describes the types of atomic elements constituting the alphabet with their type.
        The ``data'' corresponds to the type of music considered by the indicated article.
    }
    \label{tab:tokenizations_elementary}
    \\
    \headTokenizationElementary
    \endfirsthead

    \caption{
        (Continued) Overview of event-based tokenization strategies based on \textit{elementary} tokens.
    }\\
    \headTokenizationElementary
    \endhead

    ABC notation~\cite{sturm2016music} 
        & Score
        & \makecell{
            \begin{tabular}{p{2.8cm}l}
                Text alphabet &
            \end{tabular}
        } 
        & Bar patching~\cite{wu2023clamp}
        & N/A
        & Monophonic
        \\
        \midrule
    MIDI-like~\cite{oore2018time} 
        & Performance 
        & \makecell{
            \begin{tabular}{p{2.8cm}l}
                \enctype{Note-ON}{MIDI value} & \enctype{Note-OFF}{MIDI value}\\
                \enctype{Time-shift}{absolute time} & \enctype{Velocity}{integer}
            \end{tabular}
        } 
        & \makecell{
            BPE~\cite{kumar2023words,zhang2023symbolic}\\
            Unigram~\cite{kumar2023words}
        }
        & 388 
        & Piano
        \\
        \midrule
    LakhNES~\cite{donahue2019lakhnes} 
        & Performance 
        & \makecell{
            \begin{tabular}{p{2.8cm}l}
                \enctype{Note-ON/OFF-[Trk]}{MIDI value} & \enctype{Time-shift}{absolute time}
            \end{tabular}
        } 
        & --
        & 630 
        & Multi-track
        \\
        \midrule
    REMI~\cite{huang2020pop} 
        & Score 
        & \makecell{
            \begin{tabular}{p{2.8cm}l}
                \enctype{Pitch}{MIDI value} & \enctype{Duration}{music time} \\
                \enctype{Velocity}{integer} & \enctype{Chord}{class} \\
                {\texttt{<Bar>}} & \enctype{Position}{music time}
            \end{tabular}
        }
        & \makecell{
            BPE~\cite{fradet2023byte,kumar2023words,zhang2023symbolic}\\
            Unigram~\cite{kumar2023words}
        }
        & 332
        & Piano
        \\
        \midrule
    REMI+~\cite{von2022figaro} 
        & Score 
        & \makecell{
            \begin{tabular}{p{2.8cm}l}
                REMI alphabet + features: & \enctype{Instrument}{class} \\
                \enctype{Time-Signature}{class} & \enctype{Tempo}{integer} \\
            \end{tabular}
        }
        & -- 
        & N/A
        & Multi-track %
        \\
        \midrule
    \makecell{\citet{lee2022commu}\\\textit{ComMU}}
        & Score 
        & \makecell{
            \begin{tabular}{p{2.8cm}l}
                REMI alphabet + metadata: &  \\
                \enctype{Instrument}{class} & \enctype{Key}{class}\\
                \enctype{Time-Signature}{class} & \enctype{BPM}{integer} \\
                \enctype{Min/Max-velocity}{integer} & \enctype{Rhythm}{class} \\
                \enctype{Pitch-range}{class} & \enctype{Number-of-measures}{number} \\
            \end{tabular}
        }
        & -- 
        & 728
        & Multi-track %
        \\
        \midrule
    MusIAC~\cite{guo2022musiac} 
        & Score
        & \makecell{
            \begin{tabular}{p{2.8cm}l}
                REMI alphabet + control info: & \enctype{Occupation}{class} \\
                \enctype{Tensile-train}{class} & \enctype{Cloud diameter}{class}\\
                \enctype{Density}{class} & \enctype{Polyphony}{class}\\
            \end{tabular}
        }  
        & -- 
        & 360         
        & Multi-track
        \\
        \midrule
    \makecell{\citet{wu2023musemorphose}\\\textit{(MuseMorphose)}} 
        & Score 
        & \makecell{
            \begin{tabular}{p{2.8cm}l}
                \enctype{Pitch-[Trk]}{MIDI value} & \enctype{Duration-[Trk]}{music time} \\
                \enctype{Velocity-[Trk]}{integer} & \enctype{Position}{music time}\\
                {\texttt{<Bar>}} & \enctype{Tempo}{integer}\\
            \end{tabular}
        }
        & -- 
        & 3440
        & Multi-track 
        \\
        \midrule
    MultiTrack~\cite{ens2020mmm} 
        & Performance
        & \makecell{
            \begin{tabular}{p{2.8cm}l}
                {\texttt{<Start-piece>}} & {\texttt{<Start-track>/<End-track>}} \\
                {\texttt{<Start-bar>/<End-bar>}} & {\texttt{<Start-fill>/<End-fill>}} \\
                \enctype{Note-ON/OFF}{MIDI value} & \enctype{Time-shift}{absolute time}\\
                \enctype{Instrument}{class} & \enctype{Density level}{integer}
            \end{tabular}
        } 
        & -- 
        & 440 
        & Multi-track
        \\
        \midrule
    \makecell{MMR~\cite{liu2022symphony}\\\textit{(SymphonyNet)}}
        & Score 
        & \makecell{
            \begin{tabular}{p{2.8cm}l}
                {\texttt{<Start-score>/<End-score>}} & {\texttt{<Start-bar>/<End-bar>}} \\
                \enctype{Chord}{class} & {\texttt{<Change-track>}}\\
                \enctype{Position}{integer} & \enctype{Pitch}{MIDI value}\\
                \enctype{Duration}{music time} & \\
            \end{tabular}   
        }
        & BPE~\cite{liu2022symphony} 
        & N/A
        & Multi-track
        \\
        \midrule
    TSD~\cite{fradet2023byte} 
        & Performance 
        & \makecell{
            \begin{tabular}{p{2.8cm}l}
                \enctype{Pitch}{MIDI value} & \enctype{Velocity}{integer}\\
                \enctype{Duration}{absolute time} & \enctype{Time-shift}{absolute time}\\
                \enctype{Rest}{absolute time} & \enctype{Program}{class}\\
            \end{tabular}   
        }
        & BPE~\cite{fradet2023byte}
        & 249
        & Multi-track
        \\
        \midrule
    Structured~\cite{hadjeres2021piano} 
        & Performance 
        & \makecell{
            \begin{tabular}{p{2.8cm}l}
                \enctype{Pitch}{MIDI value} & \enctype{Velocity}{integer}\\
                \enctype{Duration}{absolute time} & \enctype{Time-shift}{absolute time}\\
            \end{tabular}   
        }
        & -- 
        & 428
        & Piano
        \\
        \midrule
    \citet{li2023pitchclass} 
        & Score
        & \makecell{
            \begin{tabular}{p{2.8cm}l}
                \enctype{Pitch-class}{class} & \enctype{Octave}{integer}\\
                \enctype{Bar}{integer} & \enctype{Position}{music time}\\
                \enctype{Duration}{music time} & \enctype{Velocity}{integer}\\
            \end{tabular}   
        }
        & -- 
        & N/A %
        & Monophonic
        \\
        \midrule
    \citet{chen2020automatic} 
        & Score (Tablatures) 
        & \makecell{
            \begin{tabular}{p{2.8cm}l}
                \enctype{Pitch}{MIDI value} & \enctype{Duration}{music time}\\
                \enctype{Bar}{integer} & \enctype{Position}{music time}\\
                \enctype{String}{integer} & \enctype{Fret}{integer}\\
                \enctype{Technique}{class} & \enctype{Grooving}{class}\\
                \enctype{Velocity}{integer} & \\
            \end{tabular}   
        }
        & -- 
        & 231
        & Guitar
        \\
        \midrule
    DadaGP~\cite{sarmento2021dadagp} 
        & Score (Tablatures) 
        & \makecell{
            \begin{tabular}{p{2.8cm}l}
                {\texttt{<start>/<end>}} & \enctype{Wait}{integer} \\
                \enctype{Instrument:note}{MIDI value} & \enctype{Drums:note}{MIDI value}\\
                \enctype{String}{integer} & \enctype{Fret}{integer}\\
                \enctype{Effect}{class} & \\
            \end{tabular}   
        }  
        & \makecell{
            BPE~\cite{kumar2023words}\\
            Unigram~\cite{kumar2023words}
        } 
        & 2140         
        & Guitar
        \\
    \bottomrule
    \end{longtable}
}

\footnotetext{An up-to-date and collaborative version of this table can be found at: \url{https://github.com/dinhviettoanle/survey-music-nlp\#event-based-tokenization}}  

}

\textit{{$\vcenter{\hbox{\tiny$\bullet$}}$~Alphabet --}}
In text, \textit{tokens} frequently denote words or subwords, which themselves are combinations of smaller elements - characters. In the MIR field, \textit{tokens} rather refer to the \textit{atomic} elements of the sequence that constitute what we refer to as an \textit{alphabet}.
This alphabet can be composed of a wide range of entities, such as chord labels, notes, decompositions of a note (\eg pitch, duration, etc.), or structural events such as bars. 
Thus, choosing an alphabet implies choosing a level at which to describe music and a set of attributes to represent it.

\textit{{$\vcenter{\hbox{\tiny$\bullet$}}$~Grouping strategy --}}
Atomic elements can be \textit{grouped} together to form more informative elements. 
These groupings can be established using fixed-size segmentations, statistically derived groupings, or expert-defined rules. %
In text, atomic elements (characters) are directly merged together to constitute tokens (words or subwords) leading to a vocabulary of increasing size. 
Similarly, music atomic elements can be grouped together to enrich the vocabulary with more informative tokens.

\vspace{1em}
\textbf{\emph{$\vcenter{\hbox{\tiny$\bullet$}}$ Building an \emph{alphabet} of atomic elements to encode music}} $\cdot$
A distinction between ``MIDI Score'' and ``MIDI Performance'' can be underlined~\cite{oore2018time}: the first one is a MIDI file converted from a sheet music format (\texttt{musicXML}, \texttt{kern}...) exactly following a written metrical grid, while the second one is a performance encoded into the MIDI protocol.
Scores include information such as exact timings and enharmonics, whereas performance data includes velocity and performance variations such as local tempo or dynamics.
In the following, we follow this distinction to organize existing alphabets for symbolic music tokenization.

On the one hand, \textit{performance-based} tokenization focuses on encoding music as sequences of performance events, 
nearly translating the gesture of an on-stage performer.
The MIDI-like tokenization~\cite{huang2018music} follows MIDI events from the basic MIDI protocol, including a vocabulary of 4 event types: \token{note\_on}, \token{note\_off}, \token{time\_shift}, and \token{velocity}.
This tokenization can be adapted for monophonic melodies~\cite{roberts2018hierarchical} or a polyphonic ensemble with a fixed number of instruments~\cite{donahue2019lakhnes} by having \token{note\_on/off} tokens specific to each instrument.
TSD (Time-Shift-Duration)~\cite{fradet2023byte} adapts the MIDI-like tokenization, using \token{duration} and \token{time\_shift} to replace pairs of \token{note\_on/off}
The Structured MIDI encoding~\cite{hadjeres2021piano} is similar to TSD but enforces the order of tokens describing a same event.
This avoids syntax errors in the context of live music generation and improves token sequence consistency by implicitly reducing the vocabulary size at each generation step.

Instead, \textit{score-based} tokenizations describe music as a time-structured system based on multiple discretization levels of time.
REMI (Revamped MIDI-derived events)~\cite{huang2020pop} uses a set of score-related elements to tokenize musical data, 
in particular \token{bar}, \token{position} and \token{duration} both being expressed in musical time instead of absolute timings. 
The use of such time encoding appears to bring consistency in rhythm.
Pitch encodings have also been adapted based on domain knowledge, by relying on pitch classes and octaves instead of raw MIDI numbers. This pitch encoding appears to provide better pitch distributions in both analysis~\cite{liang2020pirhdy} and generative tasks~\cite{li2023pitchclass}.
Multiple extensions of REMI have been implemented, adding additional tokens including metadata~\cite{lee2022commu}, musical features~\cite{von2022figaro,shih2022theme}, control tokens~\cite{guo2022musiac}, hand positioning for piano music~\cite{gover2022music} or track information~\cite{wu2023musemorphose}.
Before MIDI-based tokenization, early representations of music as sequences rely on score elements~\cite{conklin1995multiple}. %
This representation, called ``viewpoints'', describes relations between successive events, such as melodic contours or positions of events in a bar.
The ABC notation has also been used as a direct way of encoding monophonic scores~\cite{sturm2016music} %
where tokens are considered to be text characters.
Basic NLP models can be simply trained on these textual data for generation~\cite{sturm2016music}.

In addition, some specificities related to the instrument or the type of music data may prompt the need for adjustments to the tokenization strategy.
Tokenization strategies for guitar tablatures have been proposed for generation tasks directly in the tablature space~\cite{chen2020automatic,sarmento2021dadagp} by adding guitar-specific tokens. %
Moreover, unlike text in language, which consists of a unique stream of words, the challenge of encoding \textit{multi-track} music (\ie multi-instrument, with potentially polyphonic tracks) involves finding a way to represent simultaneous streams as a single sequence of tokens.
The representations from MMM (Multi-track Music Machine)~\cite{ens2020mmm}, MuMIDI~\cite{ren2020popmag} and the MMR (Multi-track Multi-instrument Repeatable) representation~\cite{liu2022symphony} deal with this issue by adding a token related to tracks.
However, MMR and MuMIDI interleave the different tracks to represent the multiple tracks into one sequence.
Instead, MMM concatenates all the tracks horizontally to get this single sequence.
In other words, comparing these multi-track tokenizations, MMM has a horizontal reading of the score by concatenating single-instrument tracks, 
while MMR and MuMIDI have a vertical reading of the score by firstly concatenating simultaneous measures or events from multiple tracks.

\vspace{1em}
\emph{\textbf{$\vcenter{\hbox{\tiny$\bullet$}}$ \emph{Grouping} atomic elements for shorter sequences and more informative tokens}} $\cdot$
When comparing text and music, textual sentences are often composed of hundreds of characters or around a dozen words, which is an amount of tokens that models such as Transformers can handle well. 
In contrast, musical sequences may be considerably longer %
due to various factors such as polyphony or multiple existing token types. 
To address this complexity issue, two approaches can be considered:
adapting the model mechanisms to handle this type of data (Section~\ref{sec:models})
or manipulating the representation of music in order to compress the sequence length by \textit{grouping} tokens together.

A textual n-gram~\cite[Chap. 3]{jurafsky2000speech} is a sequence of $n$ elements (characters, words, etc.) grouped together based on a fixed number of elements to constitute a token. 
N-grams have been one of the earliest representations of music borrowed from NLP~\cite{downie1999evaluating}, then improved by n-grams/skip-grams~\cite{sears2017modeling}. 
However, while grouping characters is straightforward for text data, musical n-grams can be of a diverse nature with groupings occurring at multiple levels.
Musical n-grams can be composed of note intervals or rhythm ratios~\cite{wolkowicz2008ngram}, musical descriptors~\cite{conklin1995multiple}, 
or chord n-grams to represent music through harmony~\cite{ogihara2008ngram}.
These musical n-grams also show statistical phenomena initially observed in text data representations.
Various laws such as the Heaps' law~\cite{serra2021heaps} or the Zipf's law~\cite{wolkowicz2008ngram,perotti2020emergence} can be observed with musical n-grams.
Musically-informed groupings can be derived from the musical structure of a sequence.
The CLaMP model~\cite{wu2023clamp}, which is based on the ABC notation that includes pipe characters to represent bars, considers a measure-based grouping.
Such musically-informed groupings are, however, little studied because note-level groupings are more suited as composite tokens (Section~\ref{sec:composite}), and higher-level structures, such as motifs or phrases, are often not well defined.

Finally, NLP studies have developed \textit{subword tokenization} methods~\cite{mielke2021between} where a vocabulary of subwords is statistically learned on a training corpus.
These include Byte-Pair Encoding (BPE)~\cite{gage1994new,sennrich2016neural}, WordPiece~\cite{schuster2012japanese} or UnigramLM~\cite{kudo2018subword}. 
Some of them have been adapted for music to create musical subwords as tokens.
The BPE algorithm is adapted for orchestral data~\cite{liu2022symphony} by exploiting the invariance of note order within a chord, to shorten sequence lengths.
More than a simple tool for shortening sequences, BPE has also been studied for its specific effects on musical data.
Multiple studies applied it on multiple encodings in order to examine how training Transformer models with input reduced by BPE affects both generation and analysis tasks.
Although BPE builds a more structured embedding space~\cite{fradet2023byte}, experiments studying the impact of BPE in music analysis tasks do not show a significant increase in performance~\cite{zhang2023symbolic}, unlike BPE applied to text~\cite{sennrich2016neural}.
Finally, UnigramLM subword tokenization is also specifically evaluated on music generation, applied to score-based music and guitar tablatures~\cite{kumar2023words}. 
Their findings indicate that both approaches contribute to improved data representation, enhance the structural quality of generated music, and enable the generation of longer sequences.

\paragraph{Composite tokens -- Music as sequence of combinations of multiple musical features} %
\label{sec:composite}
While sequences of elementary tokens have to introduce an artificial sequentiality by ordering musical features that describe a single event, \textit{composite tokens} encapsulate the entirety of a temporal event by combining all the musical features of this event into a single ``super-token''. 
The choice of the nature of the super-tokens, of which musical features to encapsulate into them, and of how the vector representing each super-token is constructed are the main variables in the approaches reviewed in the following. Table~\ref{tab:tokenizations_composite} describes the type of super-token and the list of features for each approach.

\afterpage{
    \newcommand{\headTokenizationComposite}{
    {\textbf{Tokenization}} & 
    {\textbf{\scriptsize\makecell[c]{Musical features}}} & 
    {\textbf{Super-token nature}} &
    {\textbf{Data}}
    \\
    \headtoprule
}

{
    \scriptsize
    \renewcommand\cellalign{cl}
    \def\arraystretch{1.3}

    \begin{longtable}{l>{\tiny}lll}
    \caption[]{
        Overview\footnotemark  of event-based tokenization strategies based on \textit{composite} tokens.
        The ``musical features'' column describes the components of the vectors considered as tokens, in terms of musical attribute.
        The ``embedded object'' denotes the manner these musical features are grouped together to form the super-token, including fixed-size vectors or based on event families.
    }
    \label{tab:tokenizations_composite}
    \\
    \headTokenizationComposite
    \endfirsthead
    
    \caption{
        (Continued) Overview of event-based tokenization strategies based on \textit{composite} tokens.
    }
    \\
    \headTokenizationComposite
    \endhead

    \citet{zhang2020adversarial} 
        & \makecell{
            \begin{tabular}{p{2.8cm}l}
                \enctype{Pitch}{integer} & \enctype{Velocity}{integer}\\
                \enctype{Program}{class} & 
            \end{tabular}
        } 
        & Homogeneous
        & Multi-track
        \\
        \midrule
    PiRhDy~\cite{liang2020pirhdy} %
        & \makecell{
            \begin{tabular}{p{2.8cm}l}
                \enctype{Chroma}{class} & \enctype{Octave}{integer}\\
                \enctype{Note-state}{class} & \enctype{Velocity}{integer} \\
                \multicolumn{2}{l}{\enctype{Inter-onset-interval}{music time}}
            \end{tabular}
        }
        & Homogeneous
        & Multi-track %
        \\
        \midrule
    \citet{zixun2021hierarchical} %
        & \makecell{
            \begin{tabular}{p{2.8cm}l}
                \enctype{Pitch}{one-hot} & \enctype{Duration}{one-hot}\\
                \enctype{Current/Next-chord}{one-hot} & \enctype{Bar}{one-hot}
                \end{tabular}
        } 
        & Homogeneous
        & Lead sheet
        \\
        \midrule
    Octuple~\cite{zeng2021musicbert} 
        & \makecell{
            \begin{tabular}{p{2.8cm}l}
                \enctype{Time-signature}{class} & \enctype{Tempo}{integer}\\
                \enctype{Bar}{integer} & \enctype{Position}{music time}\\
                \enctype{Instrument}{class} & \enctype{Pitch}{MIDI value}\\
                \enctype{Duration}{music time} & \enctype{Velocity}{integer}
            \end{tabular}
        }
        & Homogeneous
        & Multi-track
        \\
        \midrule
    \makecell{\citet{dong2023mmt}\\\textit{(MMT)}} 
        & \makecell{
            \begin{tabular}{p{2.8cm}l}
                \enctype{Type}{class} & \enctype{Beat}{integer}\\
                \enctype{Position}{music time} & \enctype{Pitch}{MIDI value}\\
                \enctype{Duration}{music time} & \enctype{Instrument}{class}
            \end{tabular} 
        } 
        & Homogeneous
        & Multi-track
        \\
        \midrule
    \makecell{\citet{dalmazzo2023chordinator}\\\textit{(Chordinator)}} %
        & \makecell{
            \begin{tabular}{p{2.8cm}l}
                \enctype{Chord-root}{class} & \enctype{Chord-nature}{class}\\
                \enctype{Chord-extensions}{class} & \enctype{Slash-chord}{boolean}\\
                \enctype{MIDI-array}{multi-hot}
            \end{tabular} 
        } 
        & Homogeneous
        & Chord sequences
        \\
        \midrule
    \makecell{\citet{wang2021musebert}\\\textit{(MuseBERT)}} 
        & \makecell{
            \begin{tabular}{p{2.8cm}l}
                \enctype{Onset}{music time} & \enctype{Pitch}{MIDI value}\\
                \enctype{Duration}{music time} & + factorized properties
            \end{tabular} 
        } 
        & Homogeneous
        & Multi-track
        \\
        \midrule
    MuMIDI~\cite{ren2020popmag} 
        & \makecell{
            \begin{tabular}{p{2.8cm}l}
                {\texttt{<Bar>}} & \enctype{Position}{music time}\\
                \enctype{Tempo}{integer} & \enctype{Track}{class}\\
                \enctype{Chord}{class} & \enctype{Pitch / Drum}{MIDI value}\\
                \enctype{Velocity}{integer} & \enctype{Duration}{music time}
            \end{tabular} 
        }
        & Family-based
        & Multi-track
        \\
        \midrule
    Compound Word~\cite{hsiao2021compound} 
        & \makecell{
            \begin{tabular}{p{2.8cm}l}
            \enctype{Family}{class} & \enctype{Time-signature}{class}\\
            \enctype{Bar}{integer} & \enctype{Beat}{music time}\\
            \enctype{Chord}{class} & \enctype{Tempo}{integer}\\
            \enctype{Pitch}{MIDI value} & \enctype{Duration}{music time}\\
            \enctype{Velocity}{integer} &
            \end{tabular} 
        }
        & Family-based
        & Piano
        \\
        \midrule
    \citet{di2021video} 
        & \makecell{
            \begin{tabular}{p{2.8cm}l}
                \enctype{Type}{class} & \enctype{Beat}{integer}\\
                \enctype{Strenth}{class} & \enctype{Density}{class}\\
                \enctype{Pitch}{MIDI value} & \enctype{Duration}{music time}\\
                \enctype{Instrument}{integer} &
            \end{tabular} 
        }
        & Family-based
        & Multi-track
        \\
        \midrule
    \citet{makris2022conditional} 
        & \makecell{
            \begin{tabular}{p{2.8cm}l}
                Encoder input: & \enctype{Onset}{number} \\
                \enctype{Group}{class} & \enctype{Type}{class}\\
                \enctype{Duration}{music time or none} & \enctype{Value}{any - depends on type}\\[.5em]
                Decoder output: & \\
                \enctype{Onset}{number} & \enctype{Drums}{integer}\\
            \end{tabular} 
        } 
        & Family-based
        & \makecell{
            Encoder:\\
            Multi-track\\[.5em]
            Decoder:\\
            Drums
        }
        \\
    \bottomrule
    \end{longtable}
}

\footnotetext{An up-to-date and collaborative version of this table can be found at: \url{https://github.com/dinhviettoanle/survey-music-nlp\#composite-tokens}}  

}

On the one hand, homogeneous super-tokens denote a representation where each super-token contains the same set of features no matter the nature of the event it describes.
The representation developed by~\citet{zixun2021hierarchical} is based on the concatenation of multiple one-hot vectors describing pitch, duration, chords, and bar. %
Octuple~\cite{zeng2021musicbert} is instead based on the embedding of 8 musical features which are concatenated to form the single vector representing a single note.
Such homogeneous representations are also used by PiRhDy~\cite{liang2020pirhdy} encoding pitch classes and octave instead of MIDI value, and MMT~\cite{dong2023mmt} for multi-track music.
Instead of vectors, MuseBERT~\cite{wang2021musebert} embeds matrices derived from a set of onset, pitch, and duration aiming at describing both musical attributes with their relations.
Beyond notes, the Chordinator model~\cite{dalmazzo2023chordinator} encodes chords described by a root, a nature, extensions, and a set of notes composing the chord.

On the other hand, methods separating events by families have been developed.
This choice is motivated by the fact that a note event is quite different from structural events such as the beginning of a bar.
For polyphonic music, MuMIDI~\cite{ren2020popmag} represents a token as a sum of the embeddings of bars, position, and tempo, with possibly note characteristics.
Similarly, Compound Word~\cite{hsiao2021compound} gathers tokens into two families: event-related or note-related and concatenates these embedded atomic elements to build the token. %
It has also been adapted for a task of drum accompaniment generation~\cite{makris2022conditional}.
This representation is also enhanced by~\citet{di2021video} in the context of video-to-music, by incorporating a token family related to rhythm, encapsulating rhythm density and strength.

\subsection{Comparing tokenization strategies}

With all these possibilities of encoding music as sequences, certain tokenization methods may demonstrate better performance on specific tasks or data than others. 
In NLP, different tokenizers, which initially aim at segmenting text, can result in different vocabularies, so that they can result in unequal performance on various tasks or languages~\cite{domingo2019much}. %
Few studies have conducted such comparisons between multiple tokenization strategies in MIR contexts.
Multiple strategies for pitch (pitch-class vs. absolute) and time grid (time resolution) encodings are compared in the context of monophonic music generation~\cite{li2023comparative}.
\citet{fradet2023impact} focus specifically on time encoding by comparing note positioning %
and duration encoding %
on generative, classification, and representation tasks. 
Beyond tokenization, a comparison between matrix, graph, and sequence representations of symbolic music is performed on various analysis tasks~\cite{zhang2023symbolic}.

Technically, the MidiTok Python package~\cite{fradet2021miditok} has been developed to provide a consistent interface for handling multiple tokenization strategies
with various tools designed to manipulate sequential symbolic music data, such as data augmentation or BPE.
Multiple other tokenizers derive from this library, including a MusicXML tokenizer~\cite{zhang2023symbolic} or a component integrated into a processing pipeline coupled with the HuggingFace library~\cite{kumar2023words}.
Similarly, Musicaiz~\cite{hernandez2023musicaiz} offers a tokenization framework, with extensive visualization, generation, and analysis frameworks for symbolic music.

\subsection{Converting music data into vectors}
\label{sec:embeddings}

The previous sections describe music encoded as sequential elements and operations that can be applied to them while keeping their high-level musical meaning. 
These elements need to be \textit{embedded} or converted into vectors so that the model can process them. 
Text, subwords, words, or documents need to be projected into a certain space in order to be processed~\cite{li2018word} leading to multiple distributional vector space models and embedding methods.

Earliest word representations simply relied on basic one-hot vectors, each with a length equivalent to the vocabulary size.
A document is represented %
by summing all these word vectors, leading to a co-occurrence counts vector, also called bag-of-words (BoW)~\cite[Chap. 4]{jurafsky2000speech}.
This representation is improved by TF-IDF (Term Frequency–Inverse Document Frequency)~\cite[Chap. 6]{jurafsky2000speech} that takes into account the total number of documents in which a word appears.
In symbolic music, such BoWs or TF-IDFs have been implemented for music similarity analysis~\cite{wolkowicz2012analysis}, mode classification in Gregorian chant~\cite{cornelissen2020mode}, or Chinese folk music clustering~\cite{liumei2021visualizing}. 
However, these approaches do not capture any sequential information and the resulting space is often sparse, preventing the ability to capture possible proximity between musical elements.
Therefore, multiple methods have been developed in the NLP field aiming at projecting words into a dense space including static and contextual embeddings.

\textit{Static embeddings} assume that each word can be encoded using the same vector
regardless of the surrounding context in which the word occurs.
Word2Vec~\cite{mikolov2013efficient} is based on a neural network that builds such static embeddings. 
This method has been adapted for music, %
implicitly leading to multiple interpretations of the definition of a musical word, including chords or musical phrases.
Multiple chord-based Word2Vec have been developed~\cite{madjiheurem2016chord2vec,huang2016chordripple}.
Such chord embeddings exhibit musical relations %
and are evaluated on downstream tasks 
like chord prediction and composer classification~\cite{lahnala2021chord}.
PitchClass2Vec~\cite{lazzari2023pitchclass2vec} embeds chords with Fasttext~\cite{bojanowski2017enriching} which relies on subwords instead of words. %
In particular, instead of embedding the whole set of pitches constituting a chord, Pitchclass2vec decomposes the chord as intervals 
in the same way as Fasttext breaks words into n-grams. %
An alternative approach considers temporal chunks of music as words.
Melody2Vec~\cite{hirai2019melody2vec} uses Word2Vec on monophonic melodies by assuming such words as musical phrases
segmented by GTTM rules~\cite{lerdahl1996generative}. 
Word2Vec has also been adapted for polyphonic music~\cite{herremans2017modeling}, by considering words as equal-length and non-overlapping slices of polyphonic music. 
Visualizing these embeddings shows a structure and organization of the space that follows the rules of tonal harmony~\cite{chuan2020context}.

Unlike static embeddings, \textit{contextual embeddings} represent a same word with different vectors depending on the context in which the word occurs
because of the polysemous nature of words. 
Although polysemy and semantics are not directly applicable in music, these contextual embeddings can be useful for symbolic music because the context in which a note appear is fundamental, for instance in functional harmony.
Technically, contextual embeddings are built concurrently with model training, such as recurrent or attention-based models (Section~\ref{sec:models}).
Yet, while analyses of learned contextual embeddings are numerous in NLP~\cite{liu2020survey},
only very few studies have specifically observed the contextual aspect of such embeddings when applied to symbolic music.
Such contextual embeddings have been analyzed from an LSTM model~\cite{garcia2020embeddings} or from BERT embeddings~\cite{han2023systematic}.
When comparing GPT-2 and BERT models, the learned embedding space from BERT is shown to be more structured than the GPT's~\cite{fradet2023byte}.
Musical context can also be defined by the relationship between simultaneous elements, extending beyond the typical temporal context encoded by classic contextual embeddings.
PiRhDy embeddings~\cite{liang2020pirhdy} encode such musical-specific context encapsulating melodic and harmonic contexts. %

\section{NLP models for symbolic music processing}
\label{sec:models}

An aspect that MIR studies have mainly borrowed from NLP is \textit{models}. 
This transfer arises primarily from the analogous temporal nature of music, which can be represented as a sequence (Section~\ref{sec:representations}), allowing its processing by NLP-based models.
Historically, in NLP, models based on recurrent cells were first implemented in the 1990s, before the breakthrough of attention models in the mid-2010s. 
MIR studies also followed these trends, adapting these models to symbolic music.

\subsection{Recurrent models}
\label{sec:recurrent_models}

Although not based on neural networks, the first sequential models applied to NLP tasks and transferred to music include Hidden Markov Models (HMM) and Conditional Random Fields (CRF) on tasks such as style classification~\cite{vercoe2001folk} or music generation~\cite{van2010music}.
In parallel, neural network-based recurrent models (RNN) have been developed and applied for symbolic music~\cite{boulangerlewandowski2012modeling}.
Improvements of basic RNNs have been at the root of several models, such as DeepBach~\cite{hadjeres2017deepbach} implementing a bidirectional LSTM~\cite{hochreiter1997long} for chorale harmonization, XiaoIceBand~\cite{zhu2018xiaoice} being based on GRU~\cite{chung2014empirical} for arrangement generation, or VirtuosoNet~\cite{jeong2019virtuosonet} implementing a hierarchical RNN~\cite{chung2017hierarchical} with an attention mechanism~\cite{bahdanau2015neural} for expressive performance generation.
These recurrent layers are implemented as part of various architectures, from variational auto-encoders with PianoTree-VAE~\cite{wang2020pianotree} to generative adversarial networks with JazzGAN~\cite{trieu2018jazzgan}.

However, from the end of the 2010s and the breakthrough of Transformer models~\cite{vaswani2017attention}, the vast majority of state-of-the-art models have now been derived from this model.
Although this survey focuses primarily on attention-based models, a thorough overview of recurrent models is available in the online material.

\subsection{Attention-based models}
\label{sec:attention_llm}

\afterpage{
    \begin{landscape}
        \newcommand{\headModelsAttention}{
    \textbf{Model}   
    & 
    & \textbf{\makecell{Base model}} 
    & \textbf{\makecell{MIR mechanism}} 
    & \textbf{Data} 
    & \textbf{\makecell{Representation}} 
    & \textbf{Tasks} 
    & \textbf{Code}
    \\
    \headtoprule
}

\renewcommand{\tableSection}[1]{\tableSectionGlobal{7}{#1}}

{   
    \scriptsize
    \renewcommand\cellalign{cl}
    \def\arraystretch{1.3}
    
    \begin{longtable}[c]{lrlllllc}
    \caption[]{
        \textit{End-to-end} Transformer-based models applied to symbolic music\footnotemark:
        such models are directly trained on specific tasks.
        Models are grouped by architecture.
        Precisions indicated in the \textit{Representation} column depict the specific adaptations brought to an initial tokenization strategy.
        The last column indicates if the code is publicly available.      
    }
    \label{tab:models_attention_end2end}
    \\
    \headModelsAttention
    \endfirsthead
    
    \caption{
        (Continued) \textit{End-to-end} Transformer-based models applied to symbolic music.
    }\\
    \headModelsAttention
    \endhead

    \tableSection{Decoder-only architecture}

    \modelauthor{Music \Tf}{\citet{huang2018music}} 
        & (2018)
        & Tf. decoder 
        & Relative attention 
        & Piano / Choral 
        & MIDI-like 
        & \makecell{Priming\\Harmonization} 
        & \availablemodel{https://github.com/jason9693/musictransformer-tensorflow2.0}
        \\
        \midrule
    \citet{chen2020automatic}
        & (2020)
        & \Tf-XL 
        & --
        & Guitar tabs 
        & \makecell{REMI-derived\\(Tablatures)}
        & \makecell{Free tabs generation} 
        & \notavailablemodel{}
        \\
        \midrule
    \modelauthor{Pop Music \Tf}{\citet{huang2020pop}} 
        & (2020)
        & \Tf-XL
        & --
        & Piano 
        & REMI 
        & \makecell{Priming\\Free generation} 
        & \availablemodel{https://github.com/YatingMusic/remi}
        \\
        \midrule
    \modelauthor{Jazz \Tf}{\citet{wu2020jazz}}
        & (2020)
        & \Tf-XL
        & --
        & Lead sheet 
        & \makecell{REMI-derived\\(Chords)} 
        & \makecell{Free generation} 
        & \availablemodel{https://github.com/slSeanWU/jazz_transformer}
        \\
        \midrule
    \modelauthor{PopMAG}{\citet{ren2020popmag}} 
        & (2020)
        & \Tf-XL %
        & --
        & Multi-track
        & MuMIDI
        & \makecell{Accompaniment generation} 
        & \notavailablemodel{}
        \\
        \midrule
    \citet{wu2020transformerxl}
        & (2020)
        & \makecell{\Tf-XL}
        & --
        & \makecell{Piano}
        & \makecell{MIDI-like-derived\\(composite tokens)}
        & \makecell{Free generation} 
        & \notavailablemodel{}
        \\
        \midrule
    \citet{di2021video}
        & (2021)
        & Tf. decoder
        & --
        & Multi-track
        & \makecell{CPWord-derived\\(Rhythm family)}
        & \makecell{Video-to-music} 
        & \availablemodel{https://github.com/wzk1015/video-bgm-generation}
        \\
        \midrule
    \citet{chang2021variable} 
        & (2021)
        & XLNet 
        & Relative bar encoding
        & Piano %
        & Compound Word
        & Infilling 
        & \availablemodel{https://github.com/reichang182/variable-length-piano-infilling}
        \\
        \midrule
    \modelauthor{Compound Word Tf.}{\citet{hsiao2021compound}} 
        & (2021)
        & Linear Tf. decoder
        & --
        & Piano 
        & Compound Word
        & \makecell{Priming\\Free generation} 
        & \availablemodel{https://github.com/YatingMusic/compound-word-transformer}
        \\
        \midrule
    \citet{sarmento2021dadagp} 
        & (2021)
        & \Tf-XL %
        & --
        & \makecell{Guitar tabs \\+ multi-track} 
        & DadaGP
        & \makecell{Metadata-conditioned gen.} 
        & \availablemodel{https://github.com/dada-bots/dadaGP}
        \\
        \midrule
    \citet{sulun2022symbolic}
        & (2022)
        & \makecell{Music \Tf} 
        & --
        & Multi-track
        & MIDI-like
        & \makecell{Emotion-conditioned gen.} 
        & \availablemodel{https://github.com/serkansulun/midi-emotion}
        \\
        \midrule
    \modelauthor{ComMU}{\citet{lee2022commu}}
        & (2022)
        & \Tf-XL
        & --
        & Multi-track 
        & REMI + metadata 
        & \makecell{Metadata-conditioned gen.\\Multi-track combination} 
        & \availablemodel{https://github.com/POZAlabs/ComMU-code}
        \\
        \midrule
    \modelauthor{SymphonyNet}{\citet{liu2022symphony}} 
        & (2022)
        & \makecell{Linear Tf.}
        & 3-D positional encoding 
        & Orchestral 
        & MMR
        & \makecell{Chord-conditioned generation\\Priming\\Free generation} 
        & \availablemodel{https://github.com/symphonynet/SymphonyNet}
        \\
        \midrule
    \citet{li2023pitchclass} 
        & (2023)
        & \Tf-XL
        & -- %
        & Lead sheet
        & \makecell{REMI-derived\\(pitch class)}
        & \makecell{Free generation} 
        & \notavailablemodel{}
        \\
        \midrule
    \modelauthor{Multitrack Music Tf.}{\citet{dong2023mmt}} 
        & (2023)
        & Tf. decoder 
        & --
        & Orchestral %
        & MMT
        & \makecell{Free generation\\Instr.-conditioned generation\\Priming} 
        & \availablemodel{https://github.com/salu133445/mmt}
        \\
        \midrule
    \modelauthor{GTR-CTRL}{\citet{sarmento2023gtr}}
        & (2023)
        & \Tf-XL 
        & --
        & \makecell{Guitar tabs \\+ multi-track} 
        & DadaGP
        & \makecell{Instr.-conditioned generation\\Genre-conditioned generation} 
        & \notavailablemodel{}
        \\
        \midrule
    \modelauthor{ShredGP}{\citet{sarmento2023shredgp}}
        & (2023)
        & \Tf-XL
        & --
        & Guitar tabs 
        & DadaGP
        & \makecell{Style-conditioned generation}
        & \notavailablemodel{}
        \\
        \midrule
    \modelauthor{Choir \Tf}{\citet{zhou2023choir}} 
        & (2023)
        & Tf. decoder
        & Relative attention
        & 4-part chorales
        & \makecell{Chord + pitch\\(event-based)}
        & \makecell{Harmonization} 
        & \availablemodel{https://github.com/Zjy0401/choir-transformer}
        \\
        \midrule
    \citet{guo2023domain}
        & (2023)
        & \makecell{Tf. encoder\\w/ custom attention} %
        & \makecell{Fundamental music embedding\\RIPO attention}
        & Monophonic 
        & FME
        & \makecell{Priming} 
        & \availablemodel{https://github.com/guozixunnicolas/fundamentalmusicembedding}
        \\
        \midrule
    \modelauthor{Compose \& Embellish}{\citet{wu2023compose}} %
        & (2023)
        & \makecell{Tf. decoder}
        & --
        & Multi-track %
        & REMI
        & \makecell{Lead sheet priming\\Accompaniment refinement} 
        & \availablemodel{https://github.com/slSeanWU/Compose_and_Embellish}
        \\
        \midrule
    \modelauthor{RHEPP-Transformer}{\citet{tang2023reconstructing}}
        & (2023)
        & \makecell{Tf. decoder} %
        & --
        & Piano
        & Octuple
        & \makecell{Expressive performance gen.} 
        & \availablemodel{https://github.com/tangjjbetsy/RHEPP-Transformer}
        \\
        \midrule
    \citet{angioni2023transformers} 
        & (2023)
        & \makecell{Tf. encoder}
        & --
        & Multi-track
        & \makecell{TSD-like}
        & \makecell{Style classification} 
        & \availablemodel{https://zenodo.org/records/7786756}
        \\
        \midrule
    \modelauthor{Chordinator}{\citet{dalmazzo2023chordinator}}
        & (2023)
        & \makecell{minGPT (no pre-training)}
        & --
        & Chords
        & \makecell{Custom chord features\\(+ MIDI array)}
        & \makecell{Chord generation} 
        & \availablemodel{https://github.com/Dazzid/theChordinator}
        \\
    \categoryrule
    \tableSection{Encoder-only architecture}
    \modelauthor{MTBert}{\citet{zhao2023forms}}
        & (2023)
        & \makecell{BERT (no pre-training)}
        & --
        & 4-part chorales
        & \makecell{Interval + duration\\(event-based)}
        & \makecell{Fugue form analysis} 
        & \notavailablemodel{}
        \\

    \categoryrule
    \tableSection{Encoder-decoder architecture}
    \modelauthor{\Tf-VAE}{\citet{jiang2020transformervae}}
        & (2020)
        & Tf. encoder-decoder
        & --
        & Monophonic
        & \makecell{Pitch + duration\\(time-slice-based)}
        & \makecell{Priming} 
        & \notavailablemodel{}
        \\
        \midrule
    \modelauthor{Harmony \Tf}{\citet{chen2021attend}} 
        & (2021)
        & Tf. encoder-decoder
        & --
        & Piano %
        & \makecell{Piano roll time-slices}
        & \makecell{Roman Numeral Analysis} 
        & \availablemodel{https://github.com/Tsung-Ping/Harmony-Transformer}
        \\
        \midrule
    \citet{makris2021generating}
        & (2021)
        & Tf. encoder-decoder
        & --
        & Lead sheet
        & \makecell{Enc.: bar features\\Dec.: chord + pitch + dur.}
        & \makecell{Emotion-conditioned gen.} 
        & \availablemodel{https://github.com/melkor169/LeadSheetGen_Valence}
        \\
        \midrule
    \citet{liutkus2021relative}
        & (2021)
        & Performer
        & Stochastic positional encoding 
        & Multi-track 
        & \makecell{REMI\\MIDI-like-derived\\(multi-track)}
        & \makecell{Free generation\\Groove continuation} 
        & \availablemodel{https://github.com/aliutkus/spe}
        \\
        \midrule
    \citet{gover2022music}
        & (2022)
        & \makecell{BART}
        & --
        & Piano
        & \makecell{REMI-derived\\(hands tokens)}
        & \makecell{Arrangement generation} 
        & \notavailablemodel{}
        \\
        \midrule
    \modelauthor{Museformer}{\citet{yu2022museformer}}
        & (2022)
        & \makecell{Tf. encoder-decoder\\w/ custom attention}
        & \makecell{Fine-/coarse-grained attention\\ Bar selection}
        & Multi-track 
        & REMI
        & \makecell{Free generation} 
        & \availablemodel{https://github.com/microsoft/muzic/tree/main/museformer}
        \\
        \midrule
    \modelauthor{Theme \Tf}{\citet{shih2022theme}}
        & (2022)
        & Tf. encoder-decoder
        & Theme-aligned pos. enc.
        & Multi-track
        & \makecell{REMI-derived\\(theme tokens)}
        & \makecell{Theme-conditioned generation} 
        & \availablemodel{https://github.com/atosystem/ThemeTransformer}
        \\
        \midrule
    \modelauthor{FIGARO}{\citet{von2022figaro}} 
        & (2022)
        & Tf. encoder-decoder 
        & --
        & Multi-track %
        & REMI+
        & \makecell{Controllable generation} 
        & \availablemodel{https://github.com/dvruette/figaro}
        \\
        \midrule
    \modelauthor{MuseMorphose}{\citet{wu2023musemorphose}} 
        & (2023)
        & Tf. enc + \Tf-XL %
        & In-attention conditioning
        & Piano %
        & \makecell{REMI-derived\\(multi-track)}
        & \makecell{Style transfer\\Controllable generation} 
        & \availablemodel{https://github.com/YatingMusic/MuseMorphose}
        \\
        \midrule
    \modelauthor{Accomontage 3}{\citet{zhao2023accomontage3}} 
        & (2023)
        & Tf. encoder-decoder
        & Instrument embedding
        & Multi-track
        & \makecell{Piano roll time-slices}
        & \makecell{Accompaniment generation} 
        & \availablemodel{https://github.com/zhaojw1998/AccoMontage-3}
        \\
        \midrule
    \modelauthor{TeleMelody}{\citet{ju2022telemelody}}
        & (2022)
        & Tf. encoder-decoder
        & --
        & Monophonic
        & \makecell{Bar + position\\+ pitch + duration\\(event-based)}
        & \makecell{Lyrics-to-melody} 
        & \availablemodel{https://github.com/microsoft/muzic/tree/main/telemelody}
        \\
        \midrule
    \modelauthor{MuseCoco}{\citet{lu2023musecoco}}
        & (2023)
        & \makecell{Text2Attr.: BERT\\Attr2Music: Linear Tf.}
        & --
        & Multi-track
        & REMI
        & Text-to-MIDI 
        & \availablemodel{https://github.com/microsoft/muzic/tree/main/musecoco}
        \\
    \categoryrule
    \tableSection{Model combinations}

    \citet{zhang2020adversarial} %
        & (2020)
        & \makecell{Generator: Tf. decoder\\Discriminator: Tf. encoder}
        & --
        & Multi-track
        & \makecell{MIDI-like-derived\\(composite tokens)} 
        & \makecell{Free generation} 
        & \notavailablemodel{}
        \\
        \midrule
    \modelauthor{\Tf-GAN}{\citet{muhamed2021symbolic}}
        & (2021)
        & \makecell{Generator: Tf.-XL\\Discriminator: BERT} %
        & -- %
        & Piano 
        & MIDI-like
        & \makecell{Free generation} 
        & \availablemodel{https://github.com/amazon-science/transformer-gan}
        \\
        \midrule
    \citet{dai2021controllable} %
        & (2021)
        & \makecell{Encoder: Tf. encoder\\Decoder: LSTM}
        & --
        & Multi-track %
        & \makecell{Pitch + rhythm\\(event-based)}
        & \makecell{Structure-conditioned gen.\\Chord conditioned gen.} 
        & \notavailablemodel{}
        \\
        \midrule
    \citet{choi2021chord} %
        & (2021)
        & \makecell{Chord enc.: Bi-LSTM\\Rhythm dec.: Tf. decoder\\Pitch dec.: Tf. decoder}
        & --
        & Lead sheet
        & \makecell{Pitch + rhythm + chord\\(time-slice-based)} 
        & \makecell{Chord-conditioned generation} 
        & \availablemodel{https://github.com/ckycky3/CMT-pytorch}
        \\
        \midrule
    \modelauthor{Bar \Tf}{\citet{qin2022bar}}
        & (2022)
        & \makecell{Bi-LSTM -\\ Tf. decoder} 
        & --
        & Lead sheet
        & \makecell{Bar + position\\+ melody + chord\\(time-slice-based)}
        & \makecell{Free generation} 
        & \notavailablemodel{}
        \\
        \midrule
    \citet{makris2022conditional} 
        & (2022)
        & \makecell{Bi-LSTM -\\ Tf. decoder}
        & --
        & \makecell{Multi-track} 
        & \makecell{CPWord-derived}
        & Drums accomp. generation 
        & \availablemodel{https://github.com/melkor169/CP_Drums_Generation}
        \\
        \midrule
    \citet{neves2022generating} %
        & (2022)
        & \makecell{Generator: Linear Tf.\\Discriminator: Linear Tf.}
        & Local prediction map
        & Piano %
        & REMI
        & \makecell{Emotion-conditioned gen.} 
        & \availablemodel{https://github.com/pneves1051/transformers_sentiment}
        \\
        \midrule
    \modelauthor{Q\&A}{\citet{zhao2023qa}} 
        & (2023)
        & \makecell{PianoTree-VAE\\Tf. decoder}
        & Instrument embedding
        & Multi-track
        & \makecell{Piano roll time-slices}
        & \makecell{Accompaniment generation} 
        & \availablemodel{https://github.com/zhaojw1998/Query-and-reArrange}
        \\
        \midrule
    \citet{duan2023melody}
        & (2023)
        & \makecell{Generator: Tf. encoder\\Discriminator: LSTM}
        & --
        & Monophonic
        & \makecell{Pitch + duration + rest\\(event-based)}
        & \makecell{Lyrics-to-melody} 
        & \notavailablemodel{}
        \\
        \midrule
    \modelauthor{Video2Music}{\citet{kang2023video2music}}
        & (2023)
        & GRU + Tf. encoder-decoder
        & --
        & Multi-track
        & MIDI-like
        & Video-to-music
        & \availablemodel{https://github.com/AMAAI-Lab/Video2Music}
        \\
        \bottomrule
    \end{longtable}
\footnotetext{An up-to-date and collaborative version of this table can be found at: \url{https://github.com/dinhviettoanle/survey-music-nlp\#end-to-end-models}}

    \begin{longtable}[c]{lrlllllc}
    \caption[]{
        \textit{Pre-trained} Transformer-based models applied to symbolic music\footnotemark:
        such models are pre-trained and then fine-tuned on downstream tasks.
    }
    \label{tab:models_attention_pretrained}
    \\
    \headModelsAttention
    \endfirsthead
    
    \caption{
        (Continued) \textit{Pre-trained} Transformer-based models applied to symbolic music.
    }\\
    \headModelsAttention
    \endhead

    \tableSection{Encoder-only architecture}
    \modelauthor{MuseBERT}{\citet{wang2021musebert}} 
        & (2021)
        & \makecell{BERT}
        & \makecell{Generalized relative pos. enc.}
        & Multi-track %
        & MuseBERT repr.
        & \makecell{ Controllable generation\\Chord analysis\\Accompaniment refinement} 
        & \availablemodel{https://github.com/ZZWaang/musebert}
        \\
        \midrule
    \modelauthor{MidiBERT-Piano}{\citet{chou2021midibert}} 
        & (2021)
        & \makecell{BERT}
        & --
        & Piano 
        & \makecell{REMI \\Compound Word}
        & \makecell{Melody extraction\\Velocity prediction\\Composer classification\\Emotion classification} 
        & \availablemodel{https://github.com/wazenmai/MIDI-BERT/}
        \\
        \midrule
    \modelauthor{MusicBERT}{\citet{zeng2021musicbert}} 
        & (2021)
        & \makecell{RoBERTa}
        & Bar-level masking
        & Multi-track*
        & Octuple
        & \makecell{Melody completion\\Accompaniment suggestion\\Genre classification\\Style classification} 
        & \availablemodel{https://github.com/microsoft/muzic/tree/main/musicbert}
        \\
        \midrule
    \modelauthor{DBTMPE}{\citet{qiu2021dbtmpe}}
        & (2021)
        & Tf. encoder
        & --
        & Multi-track
        & \makecell{Pitch combinations\\+ duration (event-based)}
        & \makecell{Style classification} 
        & \notavailablemodel{}
        \\
        \midrule
    \modelauthor{MRBERT}{\citet{li2023mrbert}} 
        & (2023)
        & \makecell{BERT}
        & \makecell{Melody/rhythm cross attention}
        & Lead sheet 
        & \makecell{Pitch + duration\\(event-based)}
        & \makecell{Free generation\\Infilling\\Chord analysis} 
        & \notavailablemodel{}
        \\
        \midrule
    \modelauthor{SoloGPBERT}{\citet{sarmento2023shredgp}}
        & (2023)
        & \makecell{BERT}
        & -- 
        & Guitar tabs 
        & DadaGP
        & \makecell{Guitar player classification}
        & \notavailablemodel{}
        \\
        \midrule
    \citet{shen2023more}
        & (2023)
        & \makecell{MidiBERT-Piano}
        & \makecell{Pre-training tasks:\\Quad-attribute masking\\Key prediction}
        & Multi-track
        & \makecell{CPWord simplified}
        & \makecell{Melody extraction\\Velocity prediction\\Composer classification\\Emotion classification} 
        & \notavailablemodel{}
        \\
        \midrule
    \modelauthor{CLaMP}{\citet{wu2023clamp}}
        & (2023)
        & \makecell{Text enc.: DistilRoBERTa\\Music enc.: BERT}
        & --
        & Lead sheet
        & \makecell{ABC notation-derived}
        & \makecell{Text-based semantic music search\\Music recommandation\\Music classification} 
        & \availablemodel{https://github.com/microsoft/muzic/tree/main/clamp}
        \\
    \categoryrule
    \tableSection{Decoder-only architecture}
    \modelauthor{LakhNES}{\citet{donahue2019lakhnes}}
        & (2019)
        & \Tf-XL 
        & --
        & Multi-track %
        & MIDI-like %
        & \makecell{Free generation} 
        & \availablemodel{https://github.com/chrisdonahue/LakhNES}
        \\
        \midrule
    \modelauthor{Musenet}{\citet{payne2019musenet}} 
        & (2019)
        & \makecell{GPT-2}
        & \makecell{Timing embedding\\Structural embedding}
        & Multi-track*
        & MIDI-like %
        & \makecell{Priming\\} 
        & \notavailablemodel{}
        \\
        \midrule
    \modelauthor{MMM}{\citet{ens2020mmm}} 
        & (2020)
        & \makecell{GPT-2}
        & --
        & Multi-track  %
        & MultiTrack repr.
        & \makecell{Free generation\\Priming\\Inpainting\\Controllable generation} 
        & \availablemodel{https://github.com/AI-Guru/MMM-JSB}
        \\
        \midrule
    \citet{angioni2023transformers} 
        & (2023)
        & \makecell{GPT-2}
        & --
        & Multi-track
        & \makecell{TSD-like}
        & \makecell{Priming} 
        & \availablemodel{https://zenodo.org/records/7786756}
        \\
        \midrule
    Zhang et. al~\cite{zhang2023drummers} %
        & (2023)
        & \makecell{GPT-3}
        & --
        & Drums
        & \makecell{Drumroll\\time-slices}
        & \makecell{Priming} 
        & \availablemodel{https://github.com/zharry29/drums-with-llm}
        \\
        \midrule
    \citet{bubeck2023sparks}
        & (2023)
        & \makecell{GPT-4}
        & --
        & Text
        & ABC notation
        & \makecell{Text-to-ABC} 
        & \notavailablemodel{}
        \\
    \categoryrule
    \tableSection{Encoder-decoder architecture}
    \modelauthor{MusIAC}{\citet{guo2022musiac}} 
        & (2022)
        & Tf. encoder-decoder
        & --
        & Multi-track %
        & REMI
        & \makecell{Infilling\\Controllable generation} 
        & \availablemodel{https://github.com/ruiguo-bio/MusIAC}
        \\
        \midrule
    \citet{li2023transformer}
        & (2023)
        & Tf. encoder-decoder
        & --
        & Lead sheet
        & \makecell{Pitch + duration\\(event-based)} %
        & \makecell{Harmony analysis\\Chord generation}  %
        & \notavailablemodel{}
        \\
        \midrule
    \citet{fu2023improve}
        & (2023)
        & \makecell{MusicBERT + Music Tf.}
        & --
        & Multi-track
        & \makecell{Octuple}
        & \makecell{Melody completion\\Accompaniment suggestion\\Melody extraction\\Emotion classification} 
        & \notavailablemodel{}
        \\
        \midrule
    \modelauthor{Multi-MMLG}{\citet{zhao2023multi}}
        & (2023)
        & \makecell{XLNet + MuseBERT}
        & --
        & \makecell{Multi-track}
        & \makecell{CPWord-derived}
        & \makecell{Melody extraction} 
        & \notavailablemodel{}
        \\
    \categoryrule
    \tableSection{Comparative studies}
    \citet{ferreira2023generating}
        & (2023)
        & \makecell{GRU, Performance-RNN \\ GPT-2 (Tf. decoder)\\Music Tf. (Tf. decoder)\\MuseNet (Tf. decoder)}
        & --
        & Piano
        & \makecell{MIDI-like}
        &  Free generation
        & \availablemodel{https://github.com/p-ferreira/generating-music-with-data}
        \\
        \midrule
    \citet{wu2023exploring}
        & (2023)
        & \makecell{BERT (Tf. encoder)\\GPT-2 (Tf. decoder)\\BART (Tf. enc.-dec.)}
        & --
        & Lead sheet
        & ABC notation
        & Text-to-ABC 
        & \availablemodel{https://github.com/sander-wood/text-to-music}
        \\
    \bottomrule

\end{longtable}
\begin{flushright}
    Tf.: \textit{Transformer} $\vert$ 
    Enc.: \textit{Encoder} $\vert$ 
    Dec.: \textit{Decoder} $\vert$ 
    Pos. enc.: \textit{Positional Encoding}
    
    (*)~These datasets are not publicly available.
\end{flushright}

\footnotetext{An up-to-date and collaborative version of this table can be found at: \url{https://github.com/dinhviettoanle/survey-music-nlp\#pre-trained-models}}  

}

    \end{landscape}
}

Attention is a mechanism proposed by~\citet{bahdanau2015neural}, initially as an improvement of RNNs. 
Vaswani et al.~\cite{vaswani2017attention} then introduced \textit{Transformers} showing that a model based solely on attention - without using any recurrent mechanism - can outperform state-of-the-art results. 
More precisely, the model is based on a \textit{self-attention} mechanism and \textit{multi-head attention} blocks.
Transformers offer two main improvements to RNNs (Section~\ref{sec:recurrent_models}).
The processing of sequences is sped up, as the entire sequence is passed through the model once and processed in parallel.
Moreover, it provides a solution to the problem of vanishing or exploding gradients that occurs with basic RNNs
and the issue of hard training with LSTMs. 
Whereby during back propagation through time, recurrent models tend to struggle in capturing long-term dependencies~\cite{noh2021analysis} between words. 
This phenomenon is also true for music generation~\cite{herremans2017functional}.

Such models have been applied to symbolic music representations,
but also in a variety of other domains, such as computer vision~\cite{dosovitskiy2020image} or audio~\cite{dong2018speech}. %
The use of Transformers has been greatly facilitated with the development of libraries, such as AllenNLP~\cite{gardner2018allennlp}, FairSeq~\cite{ott2019fairseq} or more predominantly, HuggingFace~\cite{wolf2020transformers}. 
This library offers model architectures, pre-trained models, tokenizers, and various utilities to simplify the development and deployment of NLP applications. 
As a result, numerous studies in the field of MIR have started utilizing this library, adapted for musical applications, by leveraging its tools and resources.
These include implementations of subword tokenizers, discussed in Section~\ref{sec:event_based} such as Byte-Pair encoding~\cite{sennrich2016neural} or Unigram~\cite{kudo2018subword} used by~\citet{kumar2023words} and model implementations such as BERT~\cite{devlin2018bert} used in MIDI-BERT~\cite{chou2021midibert} or GPT-2~\cite{radford2019language} used in MMM~\cite{ens2020mmm}.

In this section, we propose an overview of these Transformer-based models applied to symbolic music data seen through three technical prisms.
A first way of characterizing these models is based on their training paradigm, namely end-to-end training on specific tasks, or pre-training and fine-tuning (Section~\ref{sec:training_paradigm}). 
In a musical sense, pre-training assumes a hypothesis of a general understanding of music.
Beyond the training process, we describe various architectures that have been implemented  (Section~\ref{sec:model_architecture}). 
The model architecture, based on Transformer encoders, decoders, or combining different types of data, also assumes hypotheses on how music is processed. 
Finally, we present the enhancements of the Transformers' internal mechanism to specifically process symbolic music data (Section~\ref{sec:inner_mechanisms}).
A summary of these Transformer-based models for symbolic MIR is presented in Tables~\ref{tab:models_attention_end2end} and~\ref{tab:models_attention_pretrained}.

\subsubsection{Training paradigms: end-to-end training and pre-training}
\label{sec:training_paradigm}
Models can first be categorized by their training paradigm. 
On the one hand, end-to-end models are models trained directly for their specific task. 
On the other hand, pre-trained models, involve a pre-training of the model for a generic task followed by a fine-tuning step on one or multiple tasks and are at the heart of large language models (LLM) in NLP.
From a musical point of view, pre-trained models aim first at modelling or understanding music globally, in the same way as modelling language at a high level in NLP~\cite{zhao2023explainability}, from which specific tasks can then be derived via fine-tuning.

\paragraph{End-to-end models} %
These models are specifically trained for a particular task, most often, generative tasks.
These include Generative Adversarial Networks (GANs)~\cite{goodfellow2014generative} based on Transformers, resulting in models for multi-track generation~\cite{zhang2020adversarial}, or emotion-driven generation~\cite{neves2022generating}.
Other systems rely on Transformer-based Variational Autoencoders (VAEs)~\cite{kingma2013auto} for priming-conditioned generation~\cite{jiang2020transformervae}, chord-conditioned generation~\cite{choi2021chord}, lyrics-conditioned generation~\cite{duan2023melody} or artistic-controllable generation~\cite{von2022figaro}.
This last task is also performed in a multi-track context~\cite{lee2022commu}, with fine-grained control of the musical features at the scale of the tracks.

End-to-end models also include several data-specific models designed to process musical data beyond notes. 
The Chordinator~\cite{dalmazzo2023chordinator} model handles chord data and is based on a minGPT architecture\footnote{\url{https://github.com/karpathy/minGPT}}, without a pre-training process.
Several models are trained on guitar tablatures, for tasks such as tabs generation~\cite{chen2020automatic}, metadata-conditioned generation~\cite{sarmento2021dadagp}, style-driven generation~\cite{sarmento2023shredgp}, or instrument-conditioned generation for bands~\cite{sarmento2023gtr}.
Beyond generative tasks, a few models performing analysis tasks have been developed using this end-to-end training fashion. 
They are trained on labeled datasets, such as roman numeral-annotated datasets~\cite{chen2019harmony,chen2021attend} or style-annotated datasets~\cite{angioni2023transformers}.

\paragraph{Pre-trained models} %
In contrast with end-to-end models, pre-trained models are usually not task-specific and follow two training phases.
The model is first \textit{pre-trained} on a large corpus of data - generally unlabeled - via generic self-supervised tasks.
Once the model is pre-trained, it is \textit{fine-tuned} on a specific downstream task by being trained on a smaller task-specific labeled dataset.
This fine-tuning step is also convenient as it requires less data than the pre-training process, and takes less time to train the model instead of multiple trainings from scratch for each existing task.
While pre-training was prior to attention-based models, the latest state-of-the-art pre-trained models are now exclusively based on Transformers both in NLP and MIR.

One of the state-of-the-art pre-trained language models is BERT (Bidirectional Encoder Representations from Transformers)~\cite{devlin2018bert}.
BERT is based on a bidirectional training approach as a masked language model: a pre-training task includes masked word prediction by taking into account its left and right context.
Multiple variations of BERT applied to symbolic music have been proposed.
MuseBERT~\cite{wang2021musebert} develops a specific representation merging musical attributes and relations and processed by the attention mechanism.
MusicBERT~\cite{zeng2021musicbert} is a model designed based on RoBERTa~\cite{liu2019roberta} and improves the pre-training step by implementing a custom bar-level masking strategy instead of the original token masking.
A model combining this MusicBERT model with a Music Transformer has been evaluated on several downstream tasks, resulting in better performances~\cite{fu2023improve}.
Instrument-specific BERTs have been implemented such as SoloGPBERT~\cite{sarmento2023shredgp} for guitar tablatures, MRBERT~\cite{li2023mrbert} for lead sheets or MidiBERT-Piano~\cite{chou2021midibert} for piano.
This model is then extended beyond piano music and improved with musically meaningful pre-training tasks~\cite{shen2023more}.

GPT (Generative Pre-trained Transformer)~\cite{radford2018improving} is, instead, pre-trained through an auto-regressive task, and is more suitable for tasks involving generation.
In NLP, multiple improvements of GPT have been developed such as GPT-2~\cite{radford2019language}, GPT-3~\cite{brown2020language} and GPT-4~\cite{bubeck2023sparks}.
For symbolic music, Musenet~\cite{payne2019musenet} and MMM~\cite{ens2020mmm} are based on GPT-2 and are trained for conditioned generation.
Another approach has been implemented for drum music generation~\cite{zhang2023drummers}: music is represented as textual data which and a pre-trained textual GPT-3 is fine-tuned on this textual representation of music.

Finally, beyond GPT and BERT, %
models that integrate pre-trained components have been developed for symbolic music purposes.
LakhNES~\cite{donahue2019lakhnes} and DBTMPE~\cite{qiu2021dbtmpe} avoid the lack of data for their respective downstream tasks by being pre-trained on larger corpora and then fine-tuned for chiptune music generation or genre classification.

\subsubsection{Model architecture: Transformer encoder / decoder and multimodal models}  
\label{sec:model_architecture}
The model architecture also characterizes the existing attention-based models.
In NLP, the architecture proposed by the first Transformer model for translation~\cite{vaswani2017attention} is based on an encoder-decoder architecture. 
Afterwards, several NLP models based on either encoders~\cite{devlin2018bert}, decoders~\cite{radford2018improving}, or with modified mechanisms have been proposed.
MIR studies have therefore leveraged these existing models to adapt them for symbolic music data.
Additionally, unlike NLP models that usually handle text for both input and output, MIR experiments have been conducted with multimodal models capable of processing different types of data, in particular for tasks like text-to-symbolic music.
These multimodal models have found application in other domains such as audio processing with MusicLM~\cite{agostinelli2023musiclm} or image processing with Dall-E~\cite{ramesh2021zero}.

\paragraph{Encoder only} %
Encoders are based on a self-attention mechanism, allowing it to have knowledge of the complete sequence.
Bidirectional models, which are based on this encoder-only architecture, have led to symbolic music adaptations of BERT such as MuseBERT~\cite{wang2021musebert}, MusicBERT~\cite{zeng2021musicbert}, MidiBERT-Piano~\cite{chou2021midibert}, MRBERT~\cite{li2023mrbert}, and SoloGPBERT~\cite{sarmento2023shredgp}. 
Going further, \citet{han2023systematic} analyze the inner embeddings from BERT when trained on symbolic music and highlight the role of specific layers on the model performance.
BERT is also used as an architecture without its pre-training process by MTBert~\cite{zhao2023forms} aiming at analyzing the sections of a fugue form.
Beyond BERT, mainly characterized by its pre-training process, Transformer encoders have also been experimented with as a component of global encoder-decoder architecture, in which the encoder keeps a defined role, as detailed below.
Such a Transformer encoder is also widely used as the discriminator module in GAN-based models~\cite{zhang2020adversarial,muhamed2021symbolic,dai2021controllable}, initially developed for generation purposes.
Indeed, as most symbolic MIR studies focus on generative tasks, such encoder-only architectures are few in number.

\paragraph{Decoder only} %
In contrast with Transformer encoders, decoders implement a \emph{masked} self-attention mechanism.
Such models only have knowledge of past tokens so that they are usually implemented for auto-regressive generative tasks.
The first Music Transformer~\cite{huang2018music} is based on a decoder-only model for priming and harmonization tasks, and is then reused by~\citet{sulun2022symbolic} for emotion-conditioned generation.
Generation is tackled by the MultiTrack Music Transformer~\cite{dong2023mmt} for instrument-conditioned generation, the Choir Transformer~\cite{zhou2023choir} for 4-part harmonization, Compose \& Embellish~\cite{wu2023compose} for lead sheet and accompaniment generation, and by~\citet{tang2023reconstructing} for expressive performance reconstruction.
Decoder-only models can also be trained through a pre-training / fine-tuning process, in particular with GPT-based models, such as Musenet~\cite{payne2019musenet} or MMM~\cite{ens2020mmm}.
By comparing multiple decoder-only architectures, such pre-trained decoder-only models appear to perform better in piano generation~\cite{ferreira2023generating}.

Several models combine recurrent models with Transformer decoders.
Q\&A~\cite{zhao2023qa} combines GRU-based PianoTree-VAEs with a Transformer decoder for arrangement generation.
In the same way, \citet{choi2021chord} use a bi-LSTM model as a chord encoder, followed by Transformer decoders as pitch and rhythm generators. %
This architecture is also implemented in the Bar Transformer model~\cite{qin2022bar} for long-term structure generation, where the LSTM captures note-level dependencies and Transformer decoders capture bar-level relations.

An issue with Transformers is the quadratic complexity of the attention mechanism with respect to the sequence length.
The Linear Transformer~\cite{katharopoulos2020transformers} improves the attention mechanism with a linear complexity.
The Compound Word Transformer~\cite{hsiao2021compound} takes advantage of this computational optimization, coupled with its shorter sequence representation, for piano music generation. 
SymphonyNet~\cite{liu2022symphony} is also based on this model to address the even longer length of orchestral pieces, necessitating this lightweight attention mechanism to effectively process such data.
Another improvement of Transformers is Transformer-XL~\cite{dai2019transformerxl}, also based on auto-regressive generation, which is able
to take into account a much longer context than Transformers.
Therefore, such models have been used in several generation studies involving 
multi-track music~\cite{wu2020transformerxl,lee2022commu}, 
piano music~\cite{huang2020pop,muhamed2021symbolic,wu2023musemorphose}, 
lead sheets~\cite{wu2020jazz,li2023pitchclass}
or guitar tablatures~\cite{chen2020automatic,sarmento2021dadagp,sarmento2023gtr,sarmento2023shredgp}.
\citet{chang2021variable} implement an improved Transformer-XL, XLNet~\cite{yang2019xlnet}, a transformer-based model that can attend to past and future in the same way as BERT, while maintaining an autoregressive predicting order. This model is trained for music infilling.

\paragraph{Encoder-decoder} %
Finally, following the architecture of the vanilla Transformer, multiple models for symbolic MIR implement an encoder-decoder architecture for various tasks. %
Functional harmony analysis has been tackled by the Harmony Transformer~\cite{chen2019harmony,chen2021attend}.
The model is based on this architecture, in which the encoder has a chord segmentation role while the decoder infers the chord symbol. %

For generative purposes, such architectures are used with an encoder which analyzes musical constraints and a decoder that generates musical content.
\citet{li2023transformer} and \citet{makris2021generating} implement similar architectures, with an encoder analyzing chord (resp. chord valence) that conditions an auto-regressive decoder for a generation task.
In the Theme Transformer model~\cite{shih2022theme}, the encoder analyzes the recurrent theme, from which the decoder generates music depending on the conditions regarding the theme position within the generated content.
MusIAC~\cite{guo2022musiac} is a framework based on an encoder-decoder architecture, in which an encoder is pre-trained as a masked language model, linked with a decoder which performs an infilling task.
Multi-MMLG~\cite{zhao2023multi} is developed for a melody extraction task. 
It implements an XLNet model aiming at classifying notes as main melody or accompaniment, followed by a modified MuseBERT model that extracts secondary melodies.
In NLP, encoder-decoder models are often implemented for translation purposes~\cite{vaswani2017attention}.
\citet{gover2022music} implement BART~\cite{lewis2019bart}, an encoder-decoder architecture with learned positional embeddings, for a task analogous to language translation in the realm of music: music arrangement.
This task is also performed by Accomontage-3~\cite{zhao2023accomontage3} for multi-track music with an encoder / multiple decoders architecture.
This encoder-decoder architecture is largely used in autoencoder architectures.
The Transformer VAE~\cite{jiang2020transformervae} implements a sampling step from a latent space, from which keys and values are derived for the cross-attention mechanism.
MuseMorphose~\cite{wu2023musemorphose} and FIGARO~\cite{von2022figaro} are models based on VAEs, developed for controllable generation, which use their latent space representations as constraints.

\paragraph{Multimodal models} %
Going beyond models handling only a specific type of data, MIR systems have been developed to deal with multiple types of data such as text or video.
In symbolic MIR, studies have explored models linking text and music, including a task of lyric-to-melody
with TeleMelody~\cite{ju2022telemelody} processing musical high-level features or \citet{duan2023melody} operating at the syllable level.
Text-to-image systems %
have been gaining in popularity these last few years resulting naturally in text-to-music systems in both audio~\cite{agostinelli2023musiclm} and symbolic music.
MuseCoco~\cite{lu2023musecoco} performs this text-to-MIDI task. 
However, most text-to-symbolic-music tasks currently process an ABC notation, as this encoding is already in a textual format~\cite{wu2023exploring}.
GPT-4 is able to perform such a text-to-ABC task, among multiple other tasks~\cite{bubeck2023sparks} but struggle at modeling musical concepts such as harmony.
Finally, beyond generative tasks, CLaMP~\cite{wu2023clamp} integrates two BERT-based models -- one for text encoding and the other for music encoding -- for an analysis task, namely a tune query task based on natural language descriptions.

Multiple systems have been experimenting with symbolic music generation for video considering the use of music in videos like soundtracks in movies.
\citet{di2021video} generate music for videos that are analyzed in terms of motion speed and saliency conditioning the generated music rhythm.
\citet{kang2023video2music} add a semantic and emotion analysis of the scene, and more specifically generate chords matching these video features.

\subsubsection{Adapting attention models inner mechanisms in the context of music} 
\label{sec:inner_mechanisms}
Extensive studies have been conducted regarding the mechanisms of Transformers applied to text data, including attention and positional encoding.
When applied to symbolic music, these mechanisms may be improved to be tailored or visualized for such different data. 

Given the human intuitive aspect of visualisation, visualizing different aspects of self-attention (\eg maps, etc.) 
has been studied.
Such visualization can show differences between attention heads being more or less specialized in chords or melody~\cite{huang2018visualizing}.
Self-attention has also been studied as a source of high-level interpretations, such as music theory insights, in terms of motifs, harmony, or temporal dependencies.
Such musical objects captured by attention are numerous, including cadential passages~\cite{keller2021musical}, musical phrases or modulating sequences~\cite{jiang2020discovering}, or consonant musical intervals~\cite{dong2023mmt}.

Multiple MIR studies have also developed positional encodings and attention mechanisms customized for the specificities of music.
With the Music Transformer model~\cite{huang2018music}, a \emph{relative positional self-attention} mechanism is developed for music generation 
enabling the processing of much longer sequences.
Similarly, the \emph{stochastic positional encoding}~\cite{liutkus2021relative} aims to be compatible with linear complexity attention. %
The specificities of multi-track music inspired the SymphonyNet model to develop a \emph{3-D positional embedding}~\cite{liu2022symphony} in which the track order is permutation invariant, unlike note or measure that must remain time-dependant. 
Musically meaningful positional encodings have been developed based on notes attributes and relations~\cite{wang2021musebert}, measures~\cite{chang2021variable}, musical themes~\cite{shih2022theme}, structure and musical time~\cite{payne2019musenet}, or instruments~\cite{zhao2023qa,zhao2023accomontage3}. %

The attention mechanism has also been adapted for symbolic music.
The Museformer model~\cite{yu2022museformer} is based on a \emph{fine-grained and coarse-grained attention} aiming at reducing the complexity of the mechanism, leveraging the expected repetitive aspect of music.
The \emph{RIPO (Relative Index, Pitch and Onset) attention}~\cite{guo2023domain} is proposed with the \emph{fundamental music embedding}, relying on the structure of symbolic music built on relative onsets and pitches.
In a context of controllable style transfer, the MuseMorphose model~\cite{wu2023musemorphose} includes an \emph{in-attention conditioning} that takes into account constraints %
in the self-attention computation.
For lead sheet data, a melody/rhythm cross attention is implemented in MRBERT~\cite{li2023mrbert}, in which these two features are merged and simultaneously processed through attention.

Training strategies with musical specificities have also been developed.
Based on a GAN architecture~\cite{goodfellow2014generative}, a \emph{local prediction map}~\cite{neves2022generating} is proposed so that the discriminator also specifies which parts of the generated sequence is real or generated.
Pre-trained models, in particular masked language models, are usually pre-trained on a token prediction task from a masked sequence and a next sentence prediction task~\cite{devlin2018bert}. 
For symbolic music, MusicBERT~\cite{zeng2021musicbert} is pre-trained with a \emph{bar-level masking}: instead of masking a single token and leveraging its Octuple representation, the pre-training process masks a type of feature for all the tokens within a bar.
This masking is improved with \emph{quad-attribute masking}~\cite{shen2023more}.
These strategies avoid information leakage between tokens, as some musical features can be easily inferred from adjacent tokens.
Taking inspiration from the multi-task pre-training approach of the original BERT model, \citet{shen2023more} also propose an analogous pre-training task with next sentence prediction with \emph{key prediction}.

\section{Discussions and future directions}
\label{sec:discussion}

The previous sections outline various NLP approaches adapted to music data, resulting in %
the development of state-of-the-art tools for multiple symbolic MIR tasks.
While these results are shown to be empirically effective, it is worth taking a step back on this practice by questioning the musical appropriation of tools that have originally been thought for natural language.
Such issues can either stem from technical challenges, as NLP methods have been specifically developed and tailored for text data, or from high-level considerations, such as inherent differences between text and symbolic music.

\subsection{Technical limitations of using NLP methods for symbolic MIR}

NLP tools have been developed to specifically process text data, a type of data that remains significantly different from symbolic music, as discussed in Section~\ref{sec:discussing_nl_music}.
These methods tailored for text data may lead to technical specificities inherent to the field of NLP, which can therefore be questioned when applied to symbolic MIR.

\subparagraph{Data availability} $\cdot$
Text data differ from symbolic music data by a much wider availability.
For example, large language models such as GPT-3~\cite{brown2020language} are trained on datasets containing 300 billion tokens. %
Compared to symbolic music, 
multiple models~\cite{ens2020mmm,von2022figaro}
are trained on the LakhMIDI dataset which is composed of 175k songs, resulting in only 26M tokens using a basic MIDI-like tokenization.
Beyond the quantitative side of symbolic music datasets, there is an unavoidable bias in terms of music style diversity, as classical and pop music is much more numerous than other styles.
Moreover, while new text data are released in large amounts, contributing to extending datasets such as CommonCrawl based on publicly available text, symbolic music data is less likely to be released at this rate. 
Thus, there is a huge gap between the amount of data needed to train text models, on which Transformers are inherently efficient with such a large amount of data, and the availability of symbolic music data.

\subparagraph{Latin alphabet and musical alphabet} $\cdot$
The Latin alphabet, on which most NLP studies are based, is composed of homogeneous elements or characters. 
In contrast, musical alphabets based on the MIDI protocol are heterogeneous, consisting of multiple types of tokens, such as velocity or duration.
Therefore, musical notes are based on combinations of these atomic elements.
This combinatorial aspect is fundamental in music as two slightly different combinations can lead to radically different notes.
In substance, this is comparable to Chinese characters that can be based on different radicals, leading to entirely different meanings~\cite{wong2022introduction}. 
Such models have been developed for Chinese NLP, and take these radicals into account~\cite{tao2019radical}.

\subsection{Discussing parallels and contrasts between natural language and music}
\label{sec:discussing_nl_music}

The use of NLP methods in MIR implies that music is associated with a kind of language, which is widely debated in the musicological community.
While sharing similarities, several differences distinguish symbolic music from text, including low-level structural properties and organization, and high-level differences, especially regarding their respective function.

\subsubsection{Structural differences between text and symbolic music}
The adaptation of NLP tools for MIR is facilitated by several similarities between music representations and text  including their sequential organization.
Yet, these analogies are limited, as some aspects such as polyphony or rhythm remain inherent to music.

\subparagraph{Time dimension in language and music} $\cdot$
While speech might have a temporal dimension in terms of speech rate~\cite{wallin2001origins}, text does not explicitly encode any of these rhythmic modulations. 
In contrast, musical rhythm is based on an isochronic grid~\cite{jackendoff2009parallels} in which notes are notated with rigorous timings, in terms of onsets and durations, beyond some microtimings linked to performance embellishments or tempo changes.

\subparagraph{Simultaneity in music} $\cdot$
In music, while sequence of notes in monophonic music can be compared to words in text, polyphony adds a dimension that does not find any analogous element in text~\cite{besson2001comparison}. 
Modeling simultaneous events in a one-dimensional sequence requires approximations. 
Polyphonic music can be considered in two different ways: music can be read vertically by modeling it as a sequence of temporal events which interleaves different parts, or instead, music can be read horizontally by concatenating each part one after the other~\cite{lemstrom2007edit}.

\subparagraph{Multimodality of music} $\cdot$
Musical constitutive elements are less homogeneous than text data. 
Textual constitutive elements are of a single type: characters and possibly punctuation.
In contrast, music symbols combine structural elements (bars, position, etc.), note-related information (pitch, duration, dynamics, etc.) and global information (tempo, instrument, etc.).
Regarding computational implementations, this possibly introduces an artificial sequentiality when modeling music because multiple musical features describing one temporal event must be ordered.

\subparagraph{Segmenting text and music} $\cdot$
While whitepsaces facilitate token segmentation of text in many languages, identifying boundaries of musical motives and phrases remain subjective~\cite{lidov1997druids} or can even overlap~\cite{hentschel2021mozart}. 
In this sense, music might be more easily compared to unsegmented languages~\cite{palmer2000tokenisation}
where word segmentation can be unclear~\cite{huang2012words}.
Therefore, the application of NLP models that perform well on space-delimited languages in the context of symbolic music can be questioned.

\sepsubparagraph
\subparagraph{Musical grammar and natural language grammar} $\cdot$
While grammar is central for natural language, %
the existence of a global grammar describing music is also not unanimously accepted, even in a specific style~\cite{dempster1998grammar}.
Multiple grammars have been proposed to describe music from a general point of view, such as GTTM~\cite{lerdahl1996generative} or the implication-realization model~\cite{narmour1990analysis}.
Harmonic concepts have also been modeled as a grammar for music. 
Such harmonic rules are established by a specific musical style or era~\cite{klein2012music}: however, something which is considered ``regular'' in a style can appear as an ``irregularity'' in another style, while still being considered as music.
This absence of "rightness" in music consolidates the idea that aesthetics plays the most prominent role in music~\cite{krausz2019rightness}.
Consequently, in MIR, the evaluation systems performing generative or even analysis tasks can be delicate due to this aesthetic dimension.

\subsubsection{Functions of natural language and music}
The question of defining the function of music has been extensively studied and discussed~\cite{jackendoff2009parallels,fornas1997text,zbikowski2009music}.
Communication is central in language because it conveys ideas, thoughts, concepts or propositions. Yet, in music, communication is often considered only one of several functions ~\cite{merriam1964anthropology}.
This musical communication is often seen as serving other purposes than conveying ideas:
the concept of semantics, which is pivotal in language, is missing or at least not essential to the appreciation of music.
Music may not carry any literal meaning, or at least that cannot be compared to linguistic meaning~\cite{lerdahl1996generative}. 
Bernstein declared on this topic~\cite[p.~33]{bernstein1959joy}:%
{
    \setlength{\leftmargini}{6ex}
    \begin{quote}
        \itshape
        Music, of all the arts, stands in a special region, unlit by any star but its own, and utterly without meaning [...] except its own, a meaning in musical terms, not in terms of words.
    \end{quote}
}

Instead, music is more associated with \textit{affect} and serves as an \textit{emotional expression based on aesthetics}~\cite{jackendoff2009parallels}.  
Beyond being provoked by music, this emotional characteristic is sometimes considered as \textit{intrinsic} to the music: some compositional process can represent or symbolize an emotion~\cite{carr2004music}. 
While highly influenced by culture, Cooke illustrates this phenomenon by describing third intervals in Western music~\cite[p.~57]{cooke1959language}:%

{
    \setlength{\leftmargini}{6ex}
    \hyphenpenalty=10000
    \begin{quote}
        \itshape
        Western composers, expressing the `rightness' or happiness by means of the major third, expressed the `wrongness' of grief by means of the minor third [...].
    \end{quote}
}

This point of view that interprets musical meaning in terms of emotional descriptions is highly debatable, as these considerations often originate from cultural effects~\cite{ozaki2023cultural} or musical education.
Indeed, in both language and music, textual signs also do not inherently carry meaning. 
Instead, meaning is attributed to these signs because a particular community, from a specific era or culture, collectively establishes an agreement to associate a certain set of signs with a particular concept~\cite[p.~21]{mcclary2002feminine}.
In music, such processes are at the root of \textit{program music}~\cite{kregor2015program}.
The question of attaching meaning or semantics to music has been a subject of extensive debate for centuries and is unlikely to have a universal answer.
Returning to a technical standpoint, this epistemological debate underscores the need of carefulness when applying NLP tools for symbolic music. %

\later{
\lb{we could probably include a short discussion on the pre-training of "musical languages", and the question of what we could expect from them (perhaps in the discussion section)}
\vt{Des modèles assument qu'on puisse transférer de la connaissance d'un style instrumental à un autre ?}

\vt{
    En musique, on considère plus séparemment les genres - jazz uniquement, etc...
    
    We HAVE to borrow terms from language to talk about music
}

\vt{Music, Language and the Brain~\cite{patel2010music} should be cited somewhere...}
}

\subsection{Future directions}
\label{sec:future_directions}

NLP studies have been developed along several axes, including various aspects that may serve as research directions for symbolic MIR studies: lighter models, explainability of representations and models, and task benchmarks.

\subsubsection{Towards lighter models}
In the field of NLP, various studies have focused on developing computationally efficient yet lighter models~\cite{zhu2023surveycompression}, especially with the rise of large language models.
Such optimizations leading to lighter models are desired for multiple reasons, including reducing training or inference time, as well as energy consumption or hardware costs.
Multiple studies have explored model compression with knowledge distillation~\cite{gou2021knowledge}.
This distillation process implements a lightweight student network which is trained to reproduce a pre-trained teacher network. 
In NLP, this has led to lightweight models such as DistilBERT~\cite{sanh2020distilbert}. %
In contrast with distillation, %
pruning methods are based on altering an initial model by removing weights.
Transformers are shown to be possibly pruned by removing most of the attention heads while keeping decent performance~\cite{michel2019sixteen} and can help model explainability~\cite{voita2019analyzing}.
Finally, model design optimizations for lightweight processings %
have been developed such as 
token skipping in PoWER-BERT~\cite{goyal2020power} or sliding window attention with cache in Mistral 7B~\cite{jiang2023mistral}.
In MIR, such advances towards lighter models have begun to be tackled in the context of audio music~\cite{douwes2023energy}.

In the field of symbolic MIR, models are currently not as big as NLP models which can reach 175B parameters in the case of GPT-3~\cite{brown2020language}.
Nevertheless, there is a growing recognition of the efficacy of lighter models for symbolic music data, including the development of Compound Words~\cite{hsiao2021compound} for smaller sequences, or smaller vocabulary resulting in smaller embeddings~\cite{li2023comparative}.
These studies emphasize a promising direction for the application of lighter models in symbolic MIR research. 
This direction may involve developing light methods specifically tailored for symbolic music, featuring fewer parameters, reduced memory usage, or shorter training or inference times.
Such light models can have practical applications in real-time symbolic music generation, including improvisation where an instantaneous inference time is required.

\subsubsection{Towards more explainability}
Deep learning models are often perceived as black boxes, lacking explanations for the decisions they make. 
Several studies address the explainability aspects of NLP tools~\cite{zhao2023explainability}.
From a technical standpoint, retrieving explanations from these tools can take various forms.
Extrinsic evaluation of a model involves assessing its performance on probing tasks. %
In NLP, these probing tasks can vary in nature~\cite{conneau2018cram}, encompassing syntactic or semantic information retrieval~\cite{kim2019probing}.
In contrast, intrinsic evaluation refers to directly analyzing the inner representations occurring in the model.
In NLP, intrinsic evaluation is frequently conducted on word embeddings to assess how well a model represents words in relation to each other by examining relations like word similarity or analogies~\cite{wang2019evaluating}.
In the context of Transformers, beyond embeddings, multiple representations can be analyzed~\cite{bracsoveanu2020visualizing}, in particular attention, being %
a particularly human-interpretable mechanism.

At a low level, while text representations are most of the time based on words, music representations can be of very different nature. 
Therefore, specific representations can gain in expressiveness by incorporating more or less musical information~\cite{mckay2018jsymbolic,kermarec2022improving}. %
More recently, rationalization (\ie providing a natural language explanation of the process) based on LLMs has been explored to provide musical descriptions of symbolic music data~\cite{krol2022towards}.
Going further, providing interpretable tools that align with human behaviour can encounter challenges due to the inherent subjectivity of music. 
In the context of music composition, stylistic aspects may offer different explanations, and certain passages may only be explained by artistic effects desired by the composer~\cite{crocker1966history}.
Despite this subjectivity and artistic aspect present in music, studying the explainability of tools for symbolic music can be a way to gain a better understanding of how models process music data.
For instance, analyzing models on simple tasks such as style classification can highlight or confirm musicological characteristics in a particular style.
Similarly, with the increasing popularity of text-to-music systems, interpreting models on such tasks may reveal relations between specific words with the resulting generated content, potentially leading to questions regarding biases within the currently available datasets of symbolic music.

\subsubsection{A need for benchmarking and comparative analysis}
Benchmarks (\ie commonly accepted combinations of datasets, tasks, and evaluation metrics against which new models can be tested) are crucial for meaningful model comparisons.
The NLP community has introduced several benchmarks such as GLUE~\cite{wang2018glue} to evaluate language understanding. Other specific NLP benchmarks have also been developed, such as cross-lingual benchmarks~\cite{liang2020xglue} %
or domain-specific benchmarks~\cite{peng2019blue}.

In symbolic MIR, there is currently an apparent lack of standardized benchmarks.
Though, some symbolic music datasets are recurrently 
used as training datasets~\cite{ji2023survey},
but they rarely come with a set of evaluation tasks. %
Such standardized bundling of datasets, tasks, and evaluation metrics for symbolic music data, similar to the past MIREX challenges\footnote{\url{https://www.music-ir.org/mirex}}, may provide better frameworks to compare and evaluate models.
This question of model evaluation is fundamental.
Subjectivity is often present in music, both in analysis tasks, such as functional harmony analysis, in which annotator biases can emerge, and in generation tasks.
Evaluation of generative systems through listening tests is even more subjective~\cite{yannakis2015ratings}, but for which evaluation metrics have been proposed~\cite{wu2020jazz,kumar2023words}.
Valuable contributions regarding these benchmarking issues can be an evaluation toolkit library aiming at retrieving features from generated pieces and comparing them to those extracted from a test set.
However, this may explain the challenges in establishing such music benchmarks: the inherent subjectivity of music aesthetics restricts the possibility of "reference data", which are essential for model evaluation.

\subsubsection{Exploring further models for symbolic MIR}
Beyond improving existing MIR models, several NLP models implement mechanisms or optimizations that can be relevant to symbolic music data.
The Longformer model~\cite{beltagy2020longformer} aims to represent long documents by implementing linear complexity attention. 
Moreover, it also manages to perform well on character-level language modelling tasks.
These two characteristics are fundamental in symbolic music, as musical sequences are often longer than textual sequences. 
Additionally, unlike text where words are often considered as basic tokens, such grouping is less direct in music, so that symbolic music tasks are more similar to textual character-level tasks.
On the representation side, BERT-sentence~\cite{reimers2019sentencebert} may be relevant in the field of symbolic MIR.
This model builds embeddings for entire sentences and performs comparisons between pairs of sentences with a faster computing time.
In symbolic music, where a recurrent question concerns music segmentation, such textual sentence-derived representation holds potential relevance.
In more practical cases, pattern matching is often used in incipit search engines such as RISM\footnote{\url{https://opac.rism.info}}: an embedding-based query method can improve the tool's flexibility.

Finally, beyond NLP and the excitement in the general public for tools based on natural language generation, another trend stemming from research studies is image generation, in particular, text-to-image systems 
which are based on \emph{diffusion models}. 
Numerous recent models now integrate state-of-the-art techniques from both domains, using diffusion models coupled with Transformer blocks for controllable generation~\cite{li2023melodydiffusion,min2023polyffusion}. %
Therefore, asobserved in recent publications and preprints (Figure~\ref{fig:ismir_nlp_words}), a new trend from recent MIR studies is to adapt models initially developed for images to process music, in the same way as state-of-the-art NLP models have been adapted for symbolic music.

\later{
\vt{
    Synonymy ? Relatedness

    Questioning multilingual => Chinese
}
}

\section{Conclusion}

Symbolic music is frequently associated with natural language, drawing parallels based on structural similarities, especially in their sequential representations and numerous shared tasks. 
Consequently, the domain of Music Information Retrieval, with a specific emphasis on studies centered on symbolic music data, frequently draws inspiration from methods employed in Natural Language Processing. 
This survey organizes these NLP tools adapted for symbolic music based on two aspects: representations and models.

The process of representing text and symbolic music through sequences, referred to as tokenization, has been widely studied in the MIR field, leading to the development of various tokenization strategies. 
In contrast with text where words are often considered as basic tokens, the diversity of symbolic music tokenization strategies mainly stems from the multimodality of music, wherein each note can be described by various features.
This results in tokenizations based on time slices or musical events, incorporating technical improvements such as token grouping or composite tokens.
These representations of symbolic music are then processed by models that draw inspiration from models initially developed to process text.
Such models have been historically based on recurrent models until the breakthrough of Transformers in the field of NLP which then spread the development of several attention-based models in the field of symbolic MIR.
Nevertheless, acknowledging the particular characteristics of music in comparison with text, numerous models have incorporated music-specific mechanisms into Transformers, such as positional encoding or specialized attention mechanisms.

Despite the great performances of these models on downstream tasks such as generation or information retrieval, this usage of NLP tools - initially tailored for text data - on symbolic music can be questioned.
This includes technical issues, but also inherent epistemological differences between text and music.
These questions can therefore lead to future directions regarding this current trend, by keeping on taking inspiration from NLP advances, such as lighter, explainable models or benchmarks, to improve tools for symbolic music generation and information retrieval.


\if 0\mode
  \bibliography{references_extracted.bib} 
\else
  \printbibliography 
\fi

\end{document}